\documentclass[%
  aps,
 amsmath,amssymb,
 reprint,%
]{revtex4-2}

\usepackage{graphicx}
\usepackage{dcolumn}
\usepackage{bm}
\usepackage[draft]{changes}

\usepackage[utf8]{inputenc}
\usepackage[T1]{fontenc}
\usepackage{mathptmx}
\usepackage{etoolbox}
\usepackage{bm}

\makeatletter
\def\@email#1#2{%
 \endgroup
 \patchcmd{\titleblock@produce}
  {\frontmatter@RRAPformat}
  {\frontmatter@RRAPformat{\produce@RRAP{*#1\href{mailto:#2}{#2}}}\frontmatter@RRAPformat}
  {}{}
}%
\makeatother
\begin{document}

\title[]{Thermal characterization of suspended fine wires across continuum to \\ free-molecular gas regimes using the 3$\omega$ method}

\author{Chuyue Peng}
\author{Joshua Ginzburg}%
\author{Uri Dickman}
\author{Jacob Bair}
\author{Matthias Kuehne}
\thanks{Author to whom correspondence should be addressed: kuehne@brown.edu}
\affiliation{%
Department of Physics, Brown University, Providence, Rhode Island 02912
}%

\date{\today}

\begin{abstract}
The 3$\omega$ method is widely used to measure the thermal conductivity and the specific heat of wires and thin films. These measurements are typically performed under high vacuum conditions, which justify the use of heat transfer models that exclude thermal losses to a surrounding fluid. Here, we study the effect of thermal conduction from a joule-heated wire to a surrounding gas on pressure-dependent 3$\omega$ measurements, and show how a one-dimensional (1D) heat-transfer model may be used to reliably determine the wire's thermal properties. We derive a full analytical solution of the 1D heat-transfer equation with finite heat-transfer coefficient $h$ and validate it experimentally using a 16-µm diameter platinum wire in air across pressures from $10^{-5}$ to $10^3$~mbar. We introduce a model for heat transfer between the wire and the surrounding gas based on kinetic gas theory that accurately describes the data across continuum to free-molecular gas regimes, with $h$ varying from near-zero in high vacuum to approximately 700 W/(m$^2\cdot$K) at atmospheric pressure. We show that use of a validated $h(p)$ model allows extracting both thermal conductivity $\kappa$ and volumetric heat capacity $\rho c_p$, whereas volumetric heat capacity can be extracted even without invoking a specific $h(p)$ model. Our approach facilitates the characterization of fine wires with moderate to low thermal conductivities and may enable accurate thermal measurements of suspended wires with diameters on the nanometer scale.
\end{abstract}

\maketitle

\section{Introduction}
There is significant interest in measuring and understanding thermal transport and material properties at the micro- to nanometer scale, because of their relevance for heat management in electronics, thermoelectrics, and other applications \cite{Cahill2003,Cahill2014,Pop2010,Balandin2011,Cheng2011}. A key method for the measurement of thermal conductivity $\kappa$ and heat capacity $c_p$ at these scales is the so-called 3$\omega$ technique. This technique relies on the joule heating of an electrical resistor via the application of an electrical current with the ac component at frequency $\omega$, which gives rise to a voltage component at the third harmonic, 3$\omega$. While early work in this field was done by Corbino in the 1910s \cite{Corbino1910,Corbino1911}, it was only in the 1990s that the technique became more widely established \cite{Cahill1990,Cahill1994,moon19963omega,Lee1997}. Thermal properties are typically determined by comparing $3\omega$ voltage measurements with an appropriate heat-transfer model that captures the relevant heat-transfer pathways for a given sample configuration and experimental condition. Several variants of the technique have been put forward \cite{Bhardwaj2022} that enable the thermal characterization of various solids \cite{Cahill1990,Lu2001,Dames2005,Hou2006}, thin films \cite{Cahill1994,Lee1997,Tong2006,Dames2013,Jaffe2020}, and fluids \cite{moon19963omega,oh2008thermal,lee2009thermal,Schiffres2011,Wang2013,hamilton2024modified}.

Of particular interest is the thermal characterization of quasi-one-dimensional objects including nanowires, nanofibers, and nanotubes (hereafter all included under the term ``wire''). The geometry for a typical 3$\omega$ measurement of these is shown in Fig.~\ref{fig:1}(a) — it consists of a four-point electrical resistance measurement where an ac current is applied via the outer two electrical contacts, while the third-harmonic voltage drop across the central section is measured using the inner two contacts \cite{Yi1999,Lu2001,Dames2005,Choi2006,Hou2006,Xing2014,Xing2014b,Mishra2020,Sekimoto2023}. In this way, the influence of contact resistances on the voltage measurement is canceled out. Further, measurements are typically carried out under high vacuum to facilitate thermal characterization of the wire by simplifying the heat-transfer problem \cite{Lu2001}. An approach is missing that allows determining the wires' thermal properties even in the more general case where the wire is suspended in a fluid such as a rarified gas atmosphere. Such an approach would allow 3$\omega$ thermal characterizations of suspended fine wires in various practically relevant settings, e.g., in dilution refrigerators,  physical property measurement systems (PPMSs), or variable temperature inserts where samples are typically immersed in a thermal exchange gas atmosphere. It could also enable $in$ $situ$ characterization in industrially relevant processes with low-pressure gas environments such as chemical vapor deposition chambers or sputtering systems. Other applications may be in controlled atmosphere research, including planetary or high-altitude simulation chambers, or in the context of missions in rarified planetary atmospheres.

Herein, we present the solution of a heat-transfer equation for the 3$\omega$ method that applies to the more general case of a fine wire suspended in a gas across continuum to free-molecular regimes, which includes the high vacuum regime as a special case. We derive analytical expressions for $V_{3\omega}$ and the temperature rise along the suspended wire section. We validate our theoretical results based on measurements of a 16-µm diameter platinum (Pt) wire suspended in air at pressures between $10^{-5}$ and $10^3$~mbar. We present a method to experimentally determine the heat-transfer coefficient $h$ from the 3$\omega$ measurement, and further develop a model based on kinetic gas theory that describes heat transfer from the wire to the surrounding gas. We find our model to well describe the experimentally determined heat-transfer coefficient $h$ across the entire range of investigated gas pressures. We show how the wire's thermal conductivity $\kappa$ and volumetric heat capacity $\rho c_p$ can be extracted from the 3$\omega$ measurements, and discuss potential benefits of a finite gas pressure for the thermal characterization of fine wires with low thermal conductivity or with diameters $d$ on the nanometer scale. 

\section{Theoretical Model}

\begin{figure}
\includegraphics[width=0.49\textwidth]{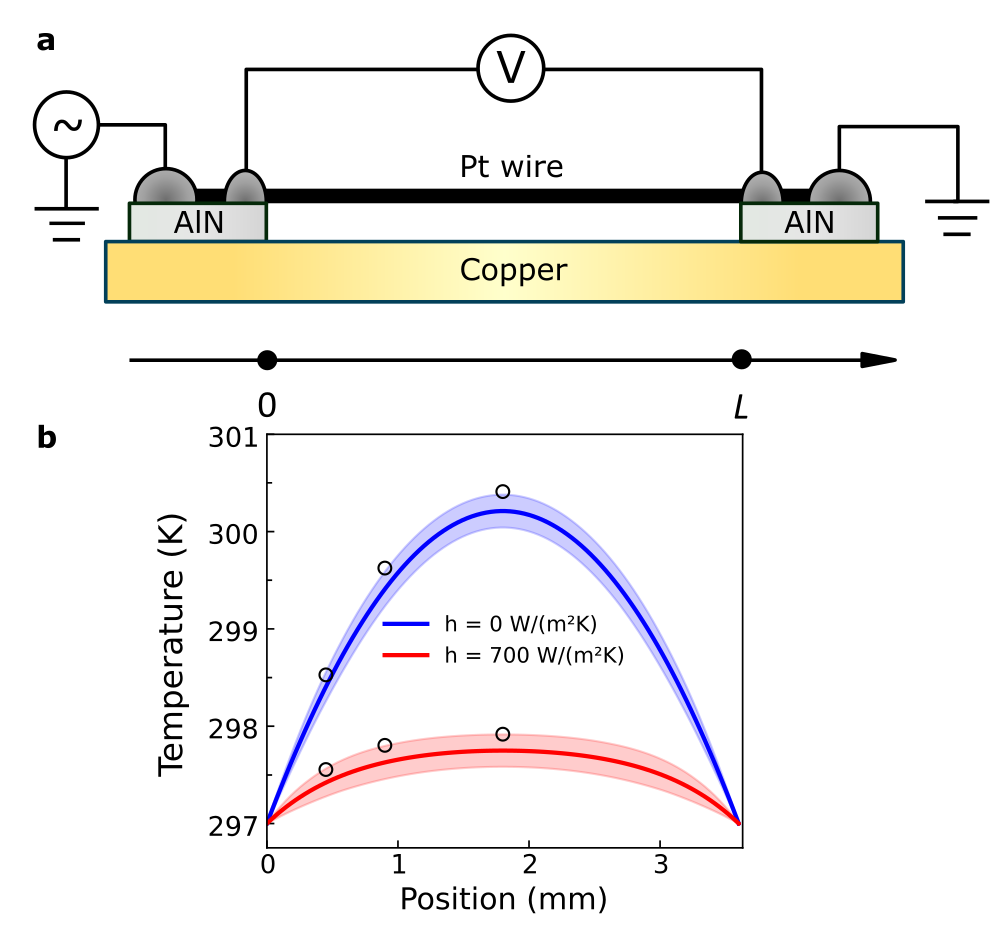}
\caption{\label{01}(a) Schematic four-terminal setup for $3\omega$ measurements of a fine wire. (b) Calculated temperature rise along the suspended wire section of length $L$ with (red) and without (blue) heat loss to a surrounding gas as quantified by the heat-transfer coefficient $h$. Solid lines illustrate the dc temperature rise, and shaded areas show the extent of ac temperature fluctuations based on Equations~(\ref{eq:temp rise with gas}) and (\ref{eq:temp rise with gasAC}). Open circles represent numerical calculations of the maximum (ac+dc) temperature rise at select positions along the wire (see Appendix B). The current used for the calculation was $I_\mathrm{rms}=7.58$~mA.}
\label{fig:1}
\end{figure}

For a free-standing wire suspended in a gas (Fig.~\ref{fig:1}(a)) and subjected to an ac electrical current applied along its length, the one-dimensional heat-transfer equation can be written as
\begin{eqnarray}
    \label{eq:1D}
    \rho c_p  \frac{\partial T(x,t)}{\partial t}  -\kappa \frac {\partial^2 T(x,t)}{\partial x^2 } +\frac{hC}{S}(T(x,t)-T_0) \nonumber\\ = \frac{I_0^2 \sin^2\omega t}{LS}[R+R'(T(x,t)-T_0)],
\end{eqnarray}
where $\rho$ is the mass density in kg/m$^3$, $c_p$ is the heat capacity in J/(kg~K), $\kappa$ is the thermal conductivity in W/(m~K), $L$ is the length of the free-standing wire section in m, $S=\pi(d/2)^2$ is the cross section of the wire in m$^2$, $C=\pi d$ is the circumference of the wire in m, $I_0$ is the peak ac current amplitude in A, $h$ is the heat-transfer coefficient in W/(m$^2$~K), $t$ is time in s, $\omega=2\pi f$ is the angular frequency in rad/s, $R$ is the electrical resistance of the wire in $\Omega$, and $R'=(\mathrm{d}R/\mathrm{d}T)_{T_0}$ is the temperature coefficient of the electrical resistance in $\Omega$/K at the substrate temperature $T_0$. This one-dimensional heat-transfer model neglects any radial temperature inhomogeneity within the wire, which is justified as long as the thermal diffusion length $\delta=\sqrt{2\alpha/\omega}$ is greater than the wire diameter $d$. Here, $\alpha = \kappa/(\rho c_p)$ is the thermal diffusivity of the wire. The suspended wire is clamped to the substrate at both sides of the free-standing wire section, which imposes the boundary condition $T(0,t) = T(L,t) = T_0$.

In Appendix A, we derive a solution for Equation~(\ref{eq:1D}) under the initial condition $T(x,-\infty)=T_0$, which can be written as $T(x,t)=T_0+\Delta T_\mathrm{dc}+\Delta T_\mathrm{ac}$, 
where $\Delta T_\mathrm{dc}$ represents the dc temperature rise, and $\Delta T_\mathrm{ac}$ represents an ac temperature variation at $2\omega$. We illustrate the effect of $h$ on the temperature profile along the suspended wire sections for parameters of interest to this work in Fig.~\ref{fig:1}(b). Results from a numerical solution of Eq.~\ref{eq:1D} (see Appendix~B) are superposed on the analytical solution at select frequencies to illustrate the accuracy of the latter. 
We further derive the rms value of the 3$\omega$ voltage:
\begin{equation}
\label{eq:rms}
    V_{3\omega,\mathrm{rms}} \approx \frac{4I_\mathrm{rms}^3 LRR'}{\pi ^4 \kappa S} \frac{1}{\sqrt{\left(1+\frac{h}{\kappa}\frac{CL^2}{\pi^2S}\right)^2+(2\omega \gamma)^2 }}.
\end{equation}
and its phase
\begin{eqnarray}
\label{eq:phase term with gas}
    \phi'=\arctan\frac{2\omega\gamma}{1+\frac{h}{\kappa}\frac{CL^2}{\pi^2S}}.
\end{eqnarray}
Here, $I_\mathrm{rms}$ is the rms value of the current and $\gamma = L^2/(\pi^2\alpha)$ is a characteristic parameter in units of time. When $h=0$, Equation (\ref{eq:rms}) recovers the vacuum case expression of \textcite{Lu2001}. Akin to their treatment of the vacuum case, we obtain the following approximation for the low-frequency (LF) limit where $\omega\gamma\rightarrow 0$:
\begin{equation}
\label{eq:v3wLF}
    V_{3\omega,\mathrm{rms,LF}} \approx \frac{4I_\mathrm{rms}^3LRR'}{\pi^4\kappa S}\frac{1}{1+\frac{h}{\kappa}\frac{CL^2}{\pi^2S}}.
\end{equation}

In the high-frequency (HF) limit, where $\omega\gamma\rightarrow \infty$, we obtain the same results as for the vacuum case:
\begin{equation}
\label{eq:v3wHF}
    V_{3\omega,\mathrm{rms,HF}} \approx \frac{I_\mathrm{rms}^3 RR'}{4\omega\rho c_p LS}.
\end{equation}

Our solution of the more general heat transfer equation (\ref{eq:1D}) makes the following predictions. First, Equation (\ref{eq:phase term with gas}) indicates that $\tan \phi'$ should be a linear function of $\omega$ with a slope of $2\gamma/(1+hC\gamma/\rho c_pS)=1/\sqrt{2}\omega_\mathrm{inflection}$. Here, 
\begin{equation}
\label{eq:omega_inflection}
    \omega_\mathrm{inflection}=\frac{1}{2\sqrt{2}\rho c_p}\left(\frac{\pi^2\kappa}{L^2}+\frac{hC}{S}\right)
\end{equation}
is the angular frequency at which the curvature of $V_{3\omega}$ changes from concave down to concave up, obtained through solving $\mathrm{d}^2 V_{3\omega}/\mathrm{d}\omega^2=0$. Second, a plot of $V_{3\omega}$ versus $I_\mathrm{rms}$ on a log-log scale should give a line with slope 3 as is established both for measurements in vacuum \cite{Lu2001,Choi2006} and in a gas atmosphere \cite{Wang2013}. Third, by combining Equations (\ref{eq:phase term with gas}) and (\ref{eq:rms}) we get
\begin{equation}
\label{eq:v3wrmsfinal}
    V_{3\omega,\mathrm{rms}}\approx\frac{2I_\mathrm{rms}^3RR'}{\pi^2\rho c_pSL\omega}\frac{1}{\sqrt{\tan^{-2}\phi'+1}},
\end{equation}
which depends on $h$ only through $\phi'$ which can be measured experimentally. An alternative way to express this is
\begin{equation}
\label{eq:v3wrmsfinal2}
    V_{3\omega,\mathrm{rms}}\approx\frac{2I_\mathrm{rms}^3RR'}{\pi^2\rho c_pSL}\frac{1}{\sqrt{2\omega_\mathrm{inflection}^2+\omega^2}}.
\end{equation}

In writing Equation~(\ref{eq:1D}), we neglect radiative heat loss that is typically assumed proportional to the circumference $C$ of the wire, its emissivity $\epsilon$, the Stefan-Boltzmann constant $\sigma=5.67\times10^{-8}$~W/(m$^2$~K$^4$), as well as the difference between the local wire temperature to the fourth power and the temperature of the environment $T_0$ to the fourth power. The small $C$ and small temperature differences on the order of a few K considered in this work significantly reduce the importance of radiation effects. According to \textcite{Lu2001}, radiative heat loss can be neglected if $16\epsilon\sigma T_0^3L^2/\pi^2\kappa d\ll 1$. Even if we assumed $\epsilon=1$ for our Pt wire, we would find $16\epsilon\sigma T_0^3L^2/\pi^2\kappa d\approx0.03$ which is indeed much smaller than 1. The actual emissivity of our annealed Pt wire is likely comparable to the one of polished Pt, i.e., $\epsilon\approx0.05$ at room temperature. We therefore ignore radiative heat loss in this work, but the interested reader may consult Appendix C and Refs.~\cite{Lu2001,Hou2006,Xing2014b} for further considerations. Our generalized 3$\omega$ method is applicable to any type of electrically conductive wire with finite $R'$ including nanowires, nanofibers, nanotubes, or bundles thereof, as long as the constraints on dimensions discussed at the end of Appendix A are respected.

\begin{figure*}
\includegraphics[width=0.99\textwidth]{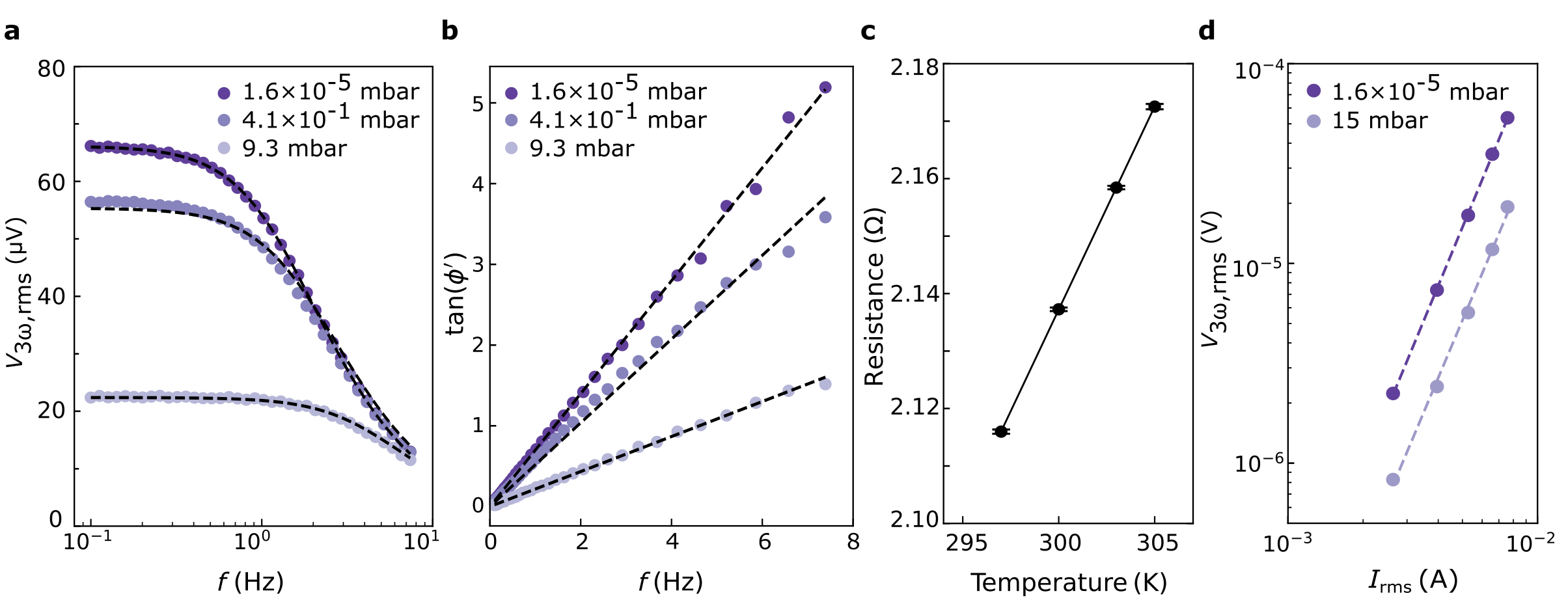}
\caption{\label{fig:02} $3\omega$ measurements of the 16-µm diameter Pt wire. (a) Root-mean-squared amplitude and (b) tangent of the phase $\phi'$ of the 3$\omega$ voltage measured at different gas pressures. 
(c) Electrical resistance $R$ of the free-standing Pt wire section measured as a function of temperature. Solid line is a linear fit of the data used to determine the temperature coefficient $R'=\mathrm{d}R/\mathrm{d}T$ of the Pt wire. (d) $V_{3\omega}$ vs $I_\mathrm{rms}$ on a log-log scale at two different pressures. All pressure-dependent data is measured at $T = 297$~K.}
\end{figure*}

\section{Experimental Results}
We conduct room-temperature $3\omega$ measurements of a 16-µm-diameter Pt wire (Goodfellow, 99.99\% pure, annealed temper) supported on an aluminum nitride (AlN) substrate (Ortech, 0.5~mm thick). We confined the experimental condition to room temperature and limited the temperature rise of the sample to below 3.5~K. The AlN substrate is an electrical insulator with high thermal conductivity ($>180$ W/(m$\cdot$K)), supported on a macroscopic copper block that acts as a heat sink and is connected to a temperature controller and thermometer. The temperature of the copper block is maintained at 297~K. Four electrical contacts are made to the Pt wire, two to each side of a 3.63-mm-long free-standing Pt wire section as schematically shown in Fig.~\ref{fig:1}(a). The four electrical contacts are made through manually placed conductive silver epoxy bonds (EPO-TEK® H20E) that also serve to ensure thermal contact of the Pt wire to the AlN substrate. The silver epoxy bonds are contacted using 32~µm diameter gold wire. The internal signal generator of a Zurich Instruments MFLI lock-in amplifier is used to drive an ac current through the Pt wire using the two outer electrical contacts, whereas the $3\omega$ voltage drop in the free-standing wire section is measured using the two inner electrical contacts. The ac current is measured simultaneously using a second demodulator of the same lock-in amplifier. Diameter and length of the free-standing Pt wire section were determined through scanning electron microscopy. The sample is loaded into a variable pressure vacuum chamber with an attached turbo pumping station and full-range pressure gauge (Pirani/cold cathode combination). The desired pressure levels are achieved by operating a diaphragm valve connected between the pump station and the chamber. For pressures higher than 10 mbar, the valve is closed after the desired pressures are reached; for pressures lower than 10~mbar, continuous pumping is required to maintain the pressure.

\begin{figure}
\includegraphics[width=0.48\textwidth]{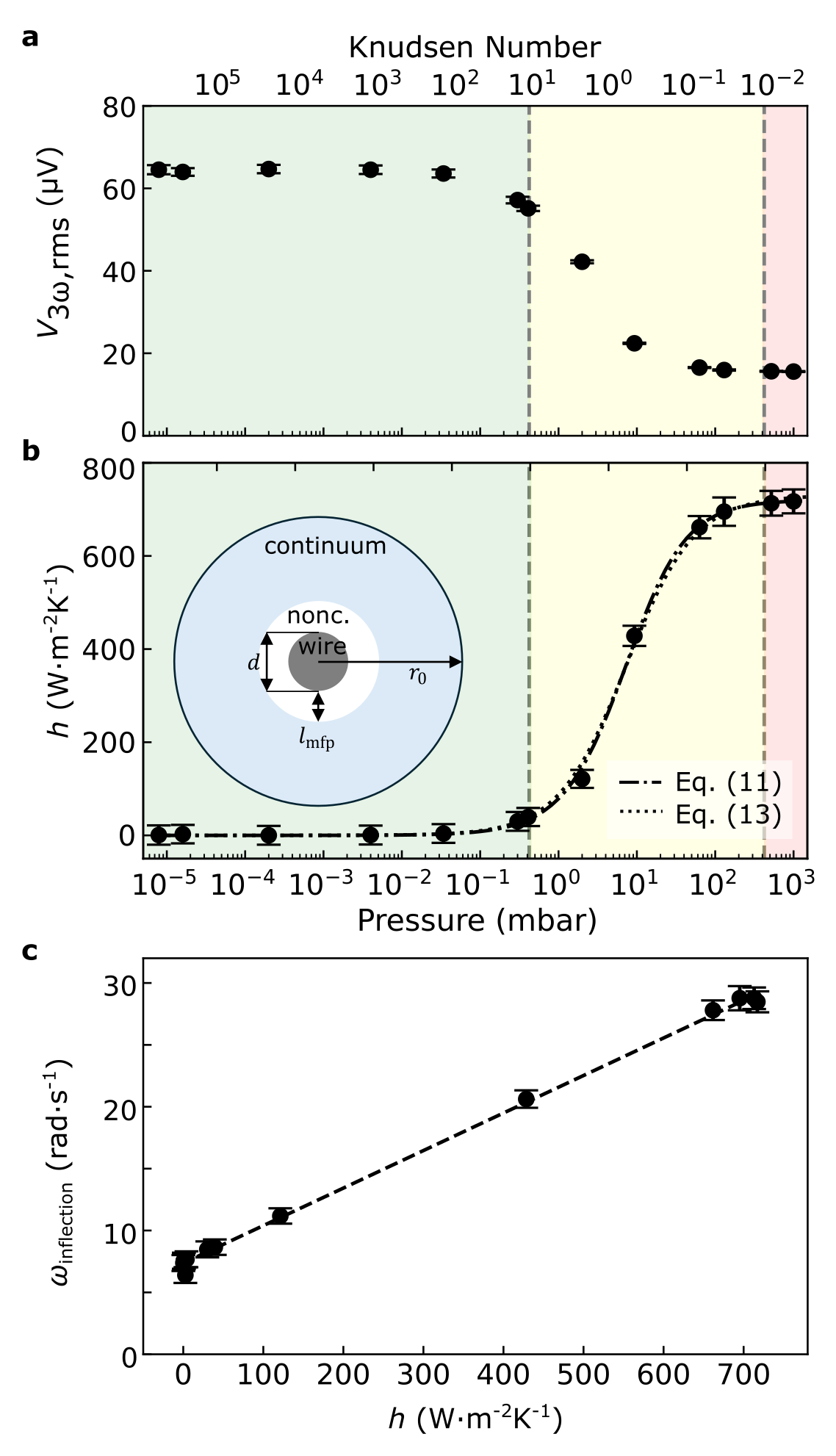}
\caption{\label{fig:03} Gas pressure-dependent 3$\omega$ measurements. (a) The $3\omega$ voltage at low frequency (0.25-0.57~Hz data averaged) is plotted as a function of gas pressure. (b) The extracted $h$ as a function of gas pressure. The dash-dotted and dotted lines are best fits of Eq.~(\ref{eq:hmodel-new}) and (\ref{eq:h}), respectively. Green, yellow, and red shaded regions indicate the free-molecular, transitional (slip and transition), and continuum regimes, respectively. The inset shows a schematic cross section of the model used to derive Eq.~(\ref{eq:hmodel-new}), subdividing space around the wire into continuum and noncontinuum (``nonc.'') regions. (c) $\omega_{\mathrm{inflection}}$ as a function of $h$. The data is obtained at 297~K.}
\end{figure}

Fig.~\ref{fig:02} shows $3\omega$ measurement results for the Pt wire. All measurements were performed with an ac current of $I_\mathrm{rms}=7.58$~mA except otherwise noted. This current induces a maximum dc temperature rise in the middle of the sample on the order of 3~K (see also Fig.~\ref{fig:1}(b)). At this constant current, different pressure levels from free molecular regime to continuum regime are achieved. The 1$\omega$, 2$\omega$, 3$\omega$ voltages and 1$\omega$ current are measured under different pressures. Fig.~\ref{fig:02}(a) is $V_{3\omega,\textrm{rms}}$ measured as a function of frequency $f=\omega/2\pi$ for three different pressures. Fig.~\ref{fig:02}(b) is the tangent of the simultaneously measured phase $\phi'$. Dashed lines in (b) are linear fits of the data. Fig.~\ref{fig:02}(c) shows the measured temperature dependence of our wire's resistance, from which we determine the thermal coefficient $R'=0.0071$~$\Omega$/K with a covariance of $5.98\times10^{-11}\Omega^2/\mathrm{K^2}$ through a linear fit of the data, shown by the solid line. Fig.~\ref{fig:02}(d) is a log-log plot of $V_{3\omega,\mathrm{rms}}$ versus $I_\textrm{rms}$ measured at $f= 1.023$~Hz. From this plot, we extract slopes of $(3.01\pm0.02)$ and $(2.97\pm0.09)$ for data measured under high vacuum conditions, $p=1.6\times 10^{-5}$~mbar, and low vacuum conditions, $p=15$~mbar, respectively. This compares favorably with the expected slope of 3 based on Equation~(\ref{eq:rms}).

We first turn to the data measured at $p=1.6\times10^{-5}$~mbar (high vacuum). At this pressure, measurements were taken at different currents, from 2.64 to 7.58~mA. In this regime, $h\approx 0$ (further justified below) and the expression for the $3\omega$ voltage simplifies to the known equation \cite{Lu2001}
\begin{eqnarray}
\label{v3w high vacuum}
    V_{3\omega} \approx -\frac{2I_0^3 LRR'}{\pi ^4 \kappa S} \frac{\sin(3\omega t -\phi)}{\sqrt{1+(2\omega \gamma)^2 }}
\end{eqnarray}
with phase
\begin{equation}
        \phi =  \arctan (2\omega \gamma).
\end{equation}
We use Equation~(\ref{v3w high vacuum}) to fit our data, yielding $\kappa=(75.8\pm5.7)$~W/(m$\cdot$K) and $\rho c_p=(3.18\pm0.41)\times10^6$~J/(m$^3$K), in good agreement with the accepted values of $\kappa=71.6$~W/(m$\cdot$K) and $\rho c_p=2.85\times10^6$~J/(m$^3$K) for Pt at room temperature \cite{Powell1966,Furukawa1974}. The stated errors represent 3 standard errors of the weighted combination of individual standard errors and observed scatter from five repeat measurements conducted under different applied currents. The downward shift observed in Fig.~\ref{fig:02}(d) when the pressure is increased to $p=15$~mbar is consistent with an increase in $h$ as predicted by Equation~(\ref{eq:v3wLF}). Consistently, the $V_{3\omega,\mathrm{rms}}$ gets increasingly suppressed as the gas pressure increases, see Fig.~\ref{fig:02}(a). According to Equation~(\ref{eq:phase term with gas}), the slope of $\tan \phi'(f)$ is also expected to decrease as the gas pressure increases, which can indeed be seen in Fig.~\ref{fig:02}(b).

Fig.~\ref{fig:03}(a) shows low-frequency $V_{3\omega,\mathrm{rms}}$ data measured from $10^{-5}-10^3$~mbar. Each data point represents an average in the low-frequency region where $V_\mathrm{3\omega,rms}$ remains constant (0.25-0.57~Hz). As the pressure decreases, the $3\omega$ voltage first increases and then reaches a plateau value of approximately equal to $64.3$~µV for $p<0.01$~mbar. This behavior suggests that $h$ approaches zero in this pressure regime, i.e., heat transfer to the surrounding gas becomes negligible compared to axial thermal conduction in the wire. We may therefore use $\kappa=(75.8\pm5.7)$~W/(m$\cdot$K) determined from the high-vacuum data to determine $h$ at higher pressures. The low-frequency $V_{3\omega,\mathrm{rms}}$ data together with the above $\kappa$ are then applied to calculate $h$ using Equation~(\ref{eq:v3wLF}). In Fig.~\ref{fig:03}(b) we plot the extracted heat-transfer coefficient $h$ obtained by this method. $h$ shows a sigmoid-type behavior, saturating at $h\approx700$~W/(m$^2\cdot$K) when approaching atmospheric pressure.

The observed behavior of $h$ can be understood based on kinetic gas theory considering thermal transport in a gas from the continuum regime at high pressures ($Kn<0.01$) to the free-molecular regime at low pressures ($Kn>10$) as determined by the value of the Knudsen number $Kn=l_\textrm{mfp}/d$ \cite{Springer1971}. For an ideal gas, the mean free path of gas molecules may be written as $l_\textrm{mfp}=k_\textrm{B}T/\sqrt{2}\pi d_\textrm{g}^2p=\zeta/p$ according to kinetic gas theory. Here, $k_\textrm{B}$ is the Boltzmann constant, and $d_\textrm{g}$ is the kinetic diameter of a gas molecule. In Appendix~D we derive an expression for $h$ based on an infinitely long-wire model that subdivides the space around the wire into two concentric regions: one close to the wire with thickness $l_\mathrm{mfp}$ in which molecules may collide with the wire but not with each other, and a second farther away from the wire in which they do collide with each other and continuum conduction holds (see inset of Fig.~\ref{fig:03}b). An expression for $h$ derived from a similar model for spherical particles has been suggested to hold for all $Kn$ \cite{Yuen1986}. Unlike in the spherical case, the logarithmic form of the temperature profile in our cylindrical geometry requires the introduction of a new length $r_0\gg d$ at which the temperature of the gas reaches $T(r_0)=T_0$. We obtain
\begin{equation}
\label{eq:hmodel-new}
    h=\frac{\kappa_\mathrm{gas}}{\frac{dl_\mathrm{mfp}\pi^2}{4\alpha_\mathrm{w}\Phi(d+2l_\mathrm{mfp})\arcsin\left(\frac{d}{d+2l_\mathrm{mfp}}\right)}+\frac{d}{2}\ln\left(\frac{2r_0}{d+2l_\mathrm{mfp}}\right)},
\end{equation}
where $\alpha_\mathrm{w}$ is the accommodation coefficient of the wire, describing the fractional extent to which molecules colliding with the wire have their mean energy adjusted to its temperature ($0\le\alpha_\mathrm{w}\le1$), and $\Phi$ is a numerical coefficient that depends on the atomicity of the gas molecules. Taking nitrogen (the main component of air) as the gas in our experiment, we use its room-temperature thermal conductivity $\kappa_\mathrm{gas}=0.026$~W/(m$\cdot$K) \cite{huber2011thermal}, and kinetic diameter $d_\textrm{g}=364$~pm going forward. $\Phi=48/95$ for such a diatomic gas \cite{Yuen1986}. Note that in the continuum regime ($Kn\rightarrow0$) Equation~(\ref{eq:hmodel-new}) simplifies to
\begin{equation}
    \label{eq:hmodel-cont}
    h=\frac{2\kappa_\mathrm{gas}}{d\ln(2r_0/d)},
\end{equation}
i.e., the high-pressure value of $h$ does not depend on $\alpha_\mathrm{w}$, but is set by $\kappa_\mathrm{gas}$ and $r_0$ instead. 

Fig.~\ref{fig:03}(b) shows a fit of our data using Eq.~(\ref{eq:hmodel-new}) which yields $\alpha_\mathrm{w}=1.09\pm 0.04$ and $r_0=(730\pm 23)$~µm. Since $\alpha_\mathrm{w}$ cannot actually exceed 1, we take our fit result as an indication that $\alpha_\mathrm{w}\approx 1$ in our experiment, which compares with a value of 0.89 reported for nitrogen on Pt \cite{Dayton1998}. Considering air as a gas mixture, we could alternatively treat $\alpha_\mathrm{w}\Phi$ as a single fit parameter (still using $\kappa_\mathrm{gas}=0.026$~W/(m$\cdot$K) and $d_\mathrm{g}=364$~pm, though), yielding $\alpha_\mathrm{w}\Phi= 0.55\pm0.02$ instead. With $\alpha_\mathrm{w}\le1$, this result would imply $\Phi\ge0.55$, which could indicate the presence of molecules with atomicity $>2$ such as water. A more refined model and better defined surface condition of the studied wire would allow for more definitive conclusions. The value of $r_0$ on the other hand compares well with the distance between the suspended Pt wire section and the copper block in our setup, which is approximately equal to $500$~µm [set by the thickness of our AlN substrate, see Fig.~\ref{fig:1}(a)]. Although our infinite wire model does not imply a specific value of $r_0$, we note that it directly compares with a textbook model for a wire in a gas-filled concentric cylinder of radius $r_0$ \cite{Dayton1998},
\begin{equation}
    \label{eq:h}
    h=\frac{\kappa_\mathrm{gas}}{\beta'\left(\frac{d}{2r_0}+1\right)l_\mathrm{mfp}+\frac{d}{2}\ln\left(\frac{2r_0}{d}\right)},
\end{equation}
in that this model also simplifies to Eq.~(\ref{eq:hmodel-cont}) in the continuum limit. This suggests that $r_0$ should indeed be comparable with the distance between the wire axis and the nearest physical surface in our experiment. In Eq.~(\ref{eq:h}), $\beta'=\frac{2-\alpha'}{\alpha'}\cdot\frac{9\gamma'-5}{2\gamma'+2}$ is a numerical value of order one, where $\alpha'$ is the accommodation coefficient assumed identical for both wire and inside surface of the concentric cylinder, and $\gamma'=c_p/c_v=1.405$ for nitrogen at room temperature \cite{Dayton1998}. Although our wire is not actually suspended inside a larger concentric cylinder, we obtain a good fit of our data using Equation~(\ref{eq:h}) with $\beta'$ and $r_0$ as fit parameters, see the dotted line in Fig.~\ref{fig:03}(b). The fit yields $r_0=(690\pm30)$~µm, which compares very well with the result obtained using Eq.~(\ref{eq:hmodel-new}) as expected. We further find $\beta'=(3.7\pm0.2)$, which implies $\alpha'=0.60\pm0.03$. This result should be interpreted with caution, as our experiment does not provide the concentric cylinder surface with accommodation coefficient $\alpha'$ required by Eq.~(\ref{eq:h}).

Fig.~\ref{fig:03}(c) plots $\omega_{\mathrm{inflection}}$ as a function of the heat-transfer coefficient $h$. We find that the data agree well with a linear model (dashed line), as predicted by Eq.~(\ref{eq:omega_inflection}). From the fitted line, we extract $\rho c_p = (2.92\pm0.11)\times10^6$~J/(m$^3$K), in good agreement with the room-temperature value for Pt, $2.85\times10^6$~J/(m$^3$K) \cite{Furukawa1974}. The value of $\kappa=(81.2\pm6.0)$~W/(m$\cdot$K) determined from the presented fit based on Eq.~(\ref{eq:omega_inflection}) is consistent with $\kappa=(75.8\pm5.7)$~W/(m$\cdot$K) that we used to calculate $h$ from our data in the first place. This $\omega_\mathrm{inflection}$ analysis provides a method for extracting $\rho c_p$ and $\kappa$ from 3$\omega$ measurements at finite gas pressure, provided a validated $h(p)$ model as derived above is employed.

\begin{figure}
\includegraphics[width=0.48\textwidth]{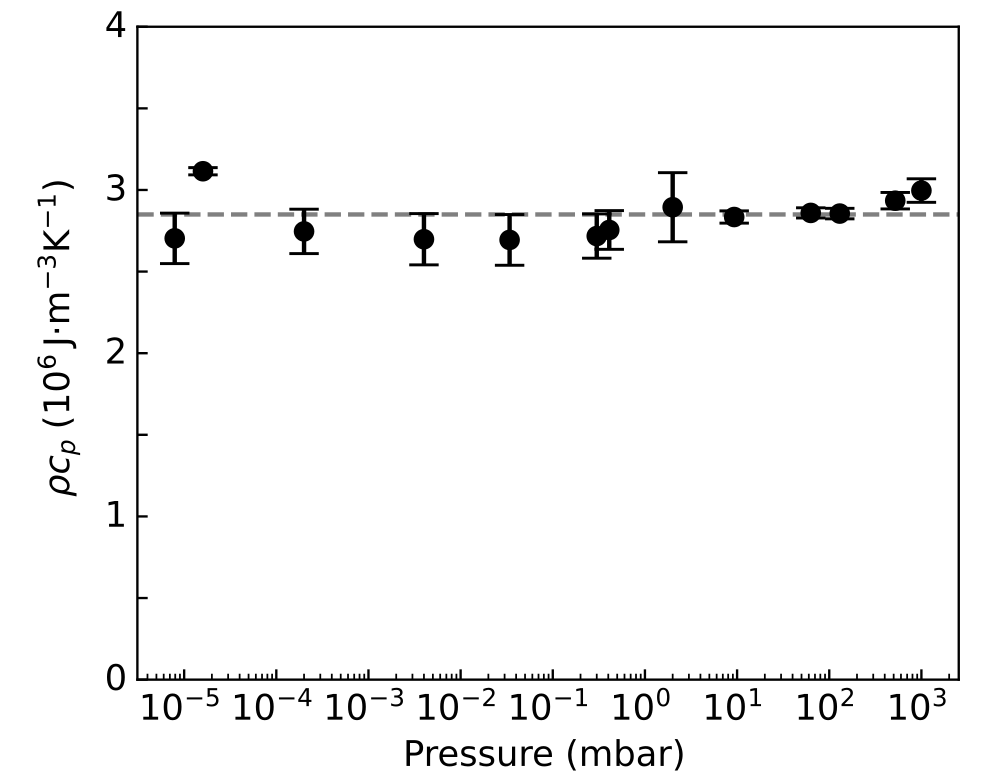}
\caption{\label{fig:04} Volumetric heat capacity of the Pt wire determined under different gas pressures. Gray dashed line is $\rho c_p=2.85\times10^6~$J/(m$^3$K) expected for Pt at 297~K \cite{Furukawa1974}.}
\end{figure}

Next, we show that even without invoking a particular $h(p)$ model, it is still possible to determine $\rho c_p$ from the measured data. Using Equation~(\ref{eq:v3wrmsfinal}), we first determine the slope of $\tan \phi'$ versus $\omega$ from a linear fit of the measured phases in Fig.~\ref{fig:02}(b) and subsequently fit the $V_{3\omega,\mathrm{rms}}$ data in Fig.~\ref{fig:02}(a) to determine $\rho c_p$. It proves difficult to reliably measure the phase $\phi'$ of the diminishing $V_{3\omega}$ signal as the frequency increases. We therefore limit the $\phi'$ analysis to frequencies below 8~Hz. This frequency range extends well beyond the low-frequency plateau of $V_{3\omega,\mathrm{rms}}$ and includes the inflection frequency $\omega_\mathrm{inflection}$ in all cases. Best fits of the $V_{3\omega,\mathrm{rms}}$ data are shown as dashed lines in Fig.~\ref{fig:02}(a), and extracted $\rho c_p$ values are shown in Fig.~\ref{fig:04}. These $\rho c_p$ values compare very well with the value expected for Pt at 297~K. Overall, the $\rho c_p$ values are consistent across 8 orders of magnitude in gas pressure with limited spread, about $\pm 10\%$. This demonstrates the feasibility of determining $\rho c_p$ of a suspended wire even at finite $h$.

\section{Discussion}

The heat-transfer coefficients $h$ extracted from our experiments inherently include interfacial thermal resistance effects. However, our analytical $h$ model does not resolve the underlying mechanisms governing thermal transport across the solid-gas interface. Instead, these complex interfacial phenomena are lumped into the accommodation coefficient $\alpha$, which serves as a phenomenological parameter that captures the aggregate effect of interface-specific thermal transport processes. Direct simulation Monte Carlo (DSMC) \cite{Bird1994,Tantos2015} or Lorentz gas approaches \cite{chen2018rough,wang2019thermal} could provide deeper mechanistic insights but go beyond the scope of this work. Both should be amenable to similarly capturing the heat transfer across continuum to free-molecular gas regimes, albeit at non-negligible computational cost. DSMC could provide molecular-level resolution of surface-gas interactions, and could serve to reveal how wire surface properties, roughness, and nonequilibrium distributions may affect thermal resistance. The Lorentz gas model could offer a complementary approach to systematically investigate how geometric factors like wire diameter and surface morphology may influence interfacial scattering patterns \cite{chen2018rough, PhysRevE.103.052135}.

\begin{figure}
    \centering
    \includegraphics[width=1\linewidth]{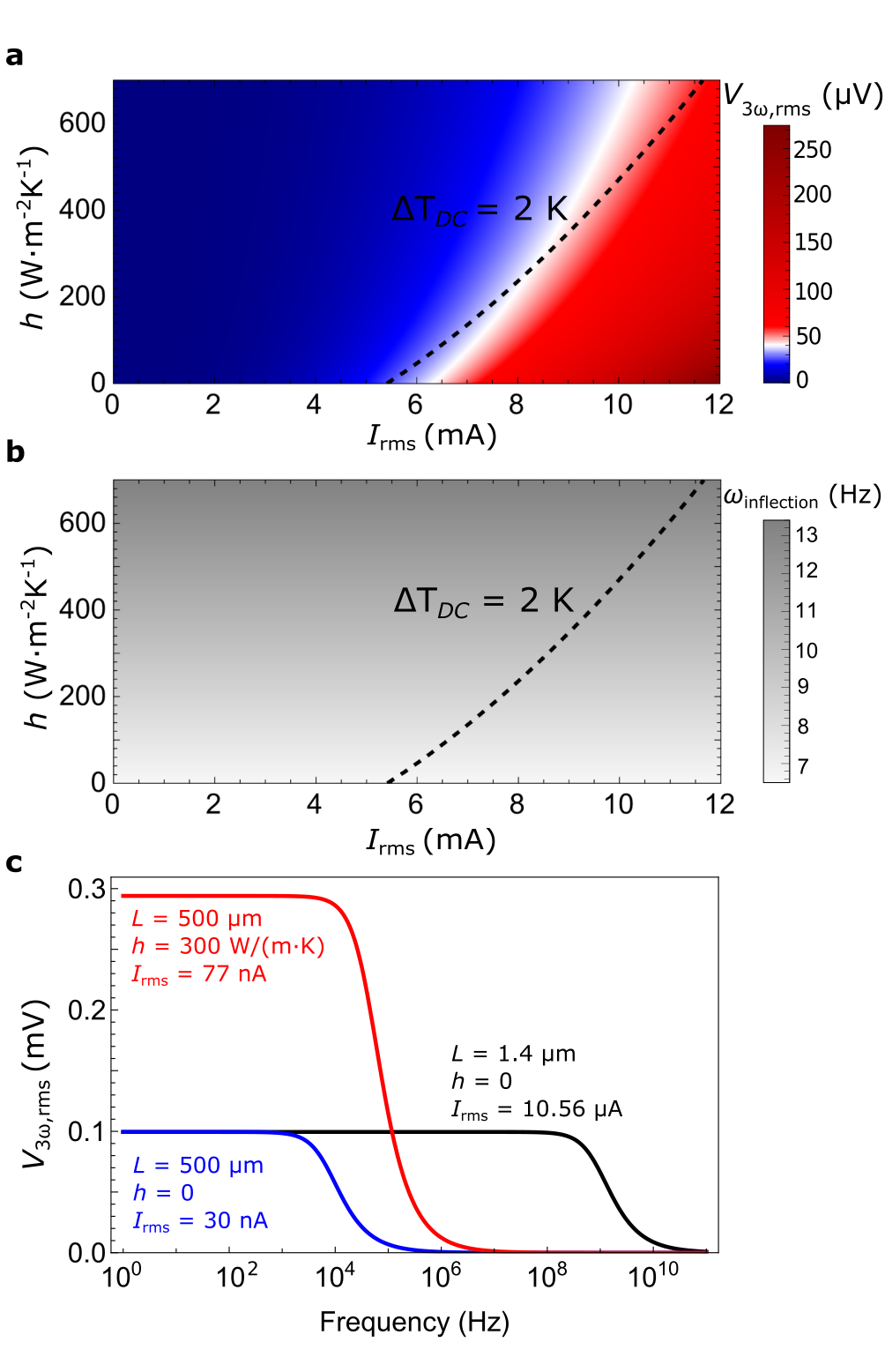}
    \caption{Benefit of finite $h$ for the thermal characterization of suspended wires. (a) Calculated $V_{3\omega,\mathrm{rms}}$ and (b) calculated $\omega_\mathrm{inflection}$ for our Pt wire. Black dashed lines in (a) and (b) are trajectories for a maximum dc temperature rise of 2~K. (c) Calculated frequency dependence of $V_\mathrm{3\omega,rms}$ of a carbon nanotube with 20-nm outer diameter.}
    \label{fig: 05}
\end{figure}

Conducting 3$\omega$ measurements at finite $h$ offers the advantage that larger $V_{3\omega,\mathrm{rms,LF}}$ signal strengths may be achieved while maintaining a fixed maximum dc temperature rise, as illustrated in Fig.~\ref{fig: 05}(a). This results essentially from the fact that $V_{3\omega,\mathrm{rms}}\propto I_\mathrm{rms}^3$ while $\Delta T_\mathrm{dc}\propto I_\mathrm{rms}^2$ (also $\Delta T_\mathrm{ac}\propto I_\mathrm{rms}^2$), as described in Equations~(\ref{eq:temp rise with gas}) and (\ref{eq:rms}). The black dashed line in Fig.~\ref{fig: 05}(a) shows a trajectory of constant maximum dc temperature rise (in the middle of the wire) of 2~K, along which $V_{3\omega,\mathrm{rms,LF}}$ increases from 24~µV at $h=0$ to 58~µV at $h=700$~W/(m$^2$K).

Such an enhancement may facilitate the characterization, e.g., of low thermal conductivity wires. Consider a fictitious wire that has the same properties as our Pt wire except for $\kappa=1$~W/(m$\cdot$K). To limit the maximum dc temperature rise to 2~K while $h=0$, we would need to reduce the applied current to $I_\mathrm{rms}=0.64$~mA. The resulting $V_{3\omega,\mathrm{rms,LF}}$ would be only $2.9$~µV. If alternatively we wanted to measure at $I_\mathrm{rms}=7.58$~mA used for most measurements of our Pt wire, we would have to shorten the wire length to $305$~µm. For this length estimation, we have used $R=0.178~\Omega$ and $R'=0.6~\mathrm{m\Omega}$/K based on the 297-K values of $R$ and $R'$ determined for our 3.63-mm-long Pt wire and assumed both constant electrical resistivity and constant temperature coefficient of the electrical resistivity. At this shorter length, $V_{3\omega,\mathrm{rms,LF}}$ would equally be $2.9$~µV. If we instead allowed for a finite $h$, e.g., $h=700$~W/(m$^2$K), we could maintain a maximum dc temperature rise of 2~K with $I_\mathrm{rms}=11$~mA applied to the 3.63-mm-long wire. The resulting $V_{3\omega,\mathrm{rms,LF}}$ would be $63$~µV, a more than an order of magnitude improvement in signal strength compared to the $h=0$ case. We recall that increasing $h$ also increases $\omega_\mathrm{inflection}$ as per Equation~(\ref{eq:omega_inflection}), and illustrate this in Fig.~\ref{fig: 05}(b).

In addition, introducing a finite $h$ can help measure $\rho c_p$ for fine wires with diameters on the nanometer scale. While shortening the length of a suspended wire may not necessarily impede the characterization of $\rho c_p$ of the wire when its diameter $d$ is on the micrometer scale, more careful consideration is required when $d$ is on the nanometer scale. To illustrate this, we consider the case of a 1.4-µm-long suspended carbon nanotube with outer and inner diameters of $d=20$~nm and $d_\mathrm{i}=10$~nm, respectively, as studied by \textcite{Choi2006}. The authors applied currents on the order of 10~\textmu A, and found $R=9.36~\mathrm{k}\Omega$ and $R'=-14.8~\Omega$/K at $T=297.5$~K. From 3$\omega$ measurements, they determined $\kappa=(300\pm20)$~W/(m$\cdot$K). However, they were not able to determine the volumetric heat capacity, which we estimate as $\rho c_p=8.5\times10^4$~J/(m$^3$K). For this estimate we use $c_p=500$~J/(kg$\cdot$K) as the room-temperature carbon-nanotube-specific heat \cite{Yi1999}, and $\rho=1700$~kg/m$^3$ as the effective-mass density considering a graphite mass density of $2260$~kg/m$^3$ smeared over the entire volume of an effective solid wire with outer diameter $d=20$~nm.

We plot the frequency dependence of $V_{3\omega,\mathrm{rms}}$ for the reported parameters and our $\rho c_p$ estimate as well as $I_\mathrm{rms}=10.55$~µA in Fig.~\ref{fig: 05}(c) (black line). The curve remains flat up to approximately 100~MHz, way beyond the upper frequency limit of $50$~kHz in the study conducted by \textcite{Choi2006}. As seen from the low-frequency approximation (Equation~(\ref{eq:v3wLF})), this regime does not provide sufficient information to extract $\rho c_p$. Information on volumetric heat capacity can only be accessed when the measurement frequency is increased above 100~MHz, which falls in a more stringent radio-frequency (rf) regime where challenges including impedance matching requirements, parasitic capacitance and inductance effects may need to be properly addressed. Moreover, at such high frequencies, the thermal diffusion length becomes smaller than the tube diameter, causing temperature gradients across the tube cross section that violate the assumption of radially uniform heating underpinning Eq.~\ref{eq:1D}.

To achieve a $V_{3\omega,\mathrm{rms}}$ curve which stops being constant at approximately 10~kHz rather than 100~MHz, the length $L$ of the nanotube must be increased. Calculated curves for $L=500$~µm are shown in Fig.~\ref{fig: 05}(c). As can be seen, to keep the dc temperature rise limited to 2~K at $h=0$, a much smaller current of $I_\mathrm{rms}=30$~nA is required (blue line). Reliably sourcing and measuring such a small current may be challenging and would likely require careful instrumentation and experiment design considerations. Allowing for $h=300$~W/(m$^2$K) instead would allow applying $I_\mathrm{rms}=77$~nA (red line in Fig.~\ref{fig: 05}(c)), which is still a small current but may be more easily sourced and measured compared to 30~nA. This would further increase $V_{3\omega,\mathrm{rms,LF}}$ by approximately a factor of 3 as can be seen in Fig.~\ref{fig: 05}(c). The associated increase in $\omega_\mathrm{inflection}$, however, may call for a balanced approach in optimizing signal strength and inflection frequency in 3$\omega$ measurements of such suspended wires with nanoscale diameters. We note that for carbon nanotubes in air and at room temperature, an upper bound of $h$ on the order of 0.1~MW/(m$^2\cdot$K) has been suggested \cite{Hu2007, Hsu2011}.

\section{Conclusions}
We conducted 3$\omega$ measurements of a 16-µm-diameter Pt wire across pressures from 10$^{-5}$ to $10^3$~mbar, validating a 1D heat-transfer model that accounts for both the wire's thermal properties ($\kappa$ and $\rho c_p$) and gas-mediated heat loss through coefficient $h$. Our analytical solution extends established vacuum-based theory to finite gas environments, with $h$ accurately described by kinetic gas theory models across continuum to free-molecular regimes. A key finding is that volumetric heat capacity $\rho c_p$ can be extracted to within $\pm 10~\%$ without requiring specific knowledge of $h(p)$, while use of validated $h(p)$ models would enable simultaneous determination of both $\kappa$ and $\rho c_p$. The finite gas-pressure approach offers practical advantages, including enhanced signal strength and reduced frequency requirements that make thermal characterization more accessible for low thermal conductivity materials and nanoscale wires where traditional vacuum methods may be limiting. This work establishes controlled gas environments as a valuable tool for 3$\omega$ thermal measurements, expanding the applicability of the technique to challenging material systems.

\begin{acknowledgments}
This material is based upon work supported by the National Science Foundation under Grant No. 2341781. J. G. acknowledges support from the Brown University Summer/Semester Projects for Research, Internship, and Teaching (SPRINT) program through both an Undergraduate Teaching and Research Awards (UTRA) fellowship as well as an Advanced Undergraduate Research Fellowship. U. D. also acknowledges support from the Brown University SPRINT program through an UTRA fellowship.

{Chuyue Peng}: Investigation, Data curation, Formal analysis, Methodology, Validation, Visualization, Writing—original draft, Writing—review \& editing. {Joshua Ginzburg}: Investigation, Formal analysis, Visualization, Writing—original draft, Writing—review \& editing. {Uri Dickman}: Investigation, Writing—review \& editing. {Jacob Bair}: Investigation, Writing—review \& editing. {Matthias Kuehne}: Conceptualization, Funding acquisition, Methodology, Project administration, Supervision, Writing—original draft, Writing—review \& editing.

\end{acknowledgments}

\section*{Data Availability Statement}
The supporting data for this article are available from Zenodo \cite{peng_2025_15586066}.

\appendix

\section{Analytical Solution of the Heat-Transfer Equation}

We solve Equation~(\ref{eq:1D}) as follows. Assuming the current to have been turned on at a time $t = -\infty$, the temperature variation from $T_0$ at any time can be calculated as the integral of the response of the temperature of the wire to the applied current at each instant of time as described by the equation
\begin{equation}
    \label{eq:impanalog}
    \Delta T(x,t) = T(x,t) -T_0 = \int_{-\infty}^tz(x,t;\tau)\,\mathrm{d}\tau.
\end{equation}
Here, $z(x,t;\tau)$ satisfies
\begin{equation}
    \label{eq:difheat}
    \frac{\partial z}{\partial t}  -\frac{\kappa}{\rho c_p}\frac {\partial^2 z}{\partial x^2 } +\frac{hC}{S\rho c_p}z - \frac{I_0^2 R'}{LS\rho c_p}z\sin^2(\omega t) = 0
\end{equation}
and is subject to the boundary conditions
\begin{align}
\begin{cases}
    \label{difcon}
    z(0,t) = 0 \\
    z(L,t)=0 \\
    z(x,\tau+0) = \frac{I_0^2 R }{LS\rho c_p}\sin^2(\omega \tau)
\end{cases}
\end{align}
Expanding $z$ as a Fourier series gives
\begin{equation}
    \label{eq:fourier}
    z(x,t;\tau) = \sum_{n = 1}^{\infty}U_n(t; \tau)\sin\frac{n\pi x}{L},
\end{equation}
which can then be substituted back into Equation~(\ref{eq:difheat}) to give
\begin{eqnarray}
    \label{eq:difheatexp}
    \sum_{n = 1}^{\infty}\left[ \frac{\mathrm{d}U_n}{\mathrm{d}t}  +\left( \frac{n^2}{\gamma}+\frac{hC}{S\rho c_p}-\frac{I_0^2 R'}{LS\rho c_p}\sin^2(\omega t)\right)U_n \right]\sin\frac{n\pi x}{L}=0\nonumber\\
\end{eqnarray}
with $\gamma = L^2/(\pi^2\alpha)$ and $\alpha = \kappa/(\rho c_p)$ as was done previously. Neglecting the $\sin^2(\omega t)$ term, since in most cases $\frac{I_0^2 R'L}{n^2\pi^2\kappa S}\ll1$ \cite{Lu2001}, Equation~(\ref{eq:difheatexp}) can be solved to give
\begin{equation}
    \label{eq:fouriercoeff}
    U_n(t;\tau) = C_n(\tau)e^{-\left( \frac{n^2}{\gamma}+\frac{hC}{S\rho c_p}\right)\left(t-\tau\right)}.
\end{equation}
Using the relation $\sum_{n = 1}^{\infty}\frac{2\left[1-(-1)^n\right]}{n\pi}\sin\frac{n\pi x}{L} = 1$ for $0 < x < L$ and the boundary conditions in~(\ref{difcon}), $C_n$ can be determined as
\begin{equation}
    \label{eq:solvedfouriercoeff}
    C_n(\tau) = \frac{2I_0^2 R\left[1-(-1)^n\right]}{n\pi LS\rho c_p}\sin^2(\omega \tau)
\end{equation}
which gives
\begin{eqnarray}
    \label{eq:solvedz}
    z(x,t;\tau) = \sum_{n = 1}^{\infty}\frac{2I_0^2 R\left[1-(-1)^n\right]}{n\pi LS\rho c_p}\sin^2(\omega \tau)\nonumber\\ \times\ e^{-\left( \frac{n^2}{\gamma}+\frac{hC}{S\rho c_p}\right)\left(t-\tau\right)} \sin\frac{n\pi x}{L}
\end{eqnarray}
when substituted back into Equation~(\ref{eq:fourier}). Finally, substituting Equation~(\ref{eq:solvedz}) into Equation~(\ref{eq:impanalog}) give the expression for the temperature variation
\begin{eqnarray}
    \label{eq:variation}
    \Delta T(x,t) = \frac{2I_0^2 RL}{\pi^3\kappa S}\sum_{n = 1}^{\infty}\frac{\left[1-(-1)^n\right]}{2n^3}\sin\left(\frac{n\pi x}{L}\right)\nonumber\\ \times\left( \frac{1}{1+\frac{h'\gamma}{n^2}}-\frac{\frac{1}{2}e^{-2i\omega t}}{1+\frac{h'\gamma}{n^2}-\frac{2i\omega \gamma}{n^2}}-\frac{\frac{1}{2}e^{2i\omega t}}{1+\frac{h'\gamma}{n^2}+\frac{2i\omega\gamma}{n^2}}\right)
\end{eqnarray}
where $h' = \frac{hC}{\rho c_p S}$. $\Delta T_\mathrm{ac}(x,t)$ corresponds to the terms in Equation~(\ref{eq:variation}) with dependence on $\omega t$, which could be written as:
\begin{align}
\label{eq:temp rise with gasAC}
    &\Delta T_\mathrm{ac}(x,t) = \frac{2I_0^2 RL}{\pi^3 \kappa S}\sum_{n = 1}^{\infty}\frac{-1+(-1)^n}{2n^3}\sin\frac{n\pi x}{L}\nonumber\\&\times\bigg[\frac{\frac{1}{2}\exp(-2i\omega t)}{1+\frac{1}{n^2}\frac{h}{\kappa}\frac{CL^2}{\pi^2S}-\frac{2i\omega\gamma}{n^2}}+\frac{\frac{1}{2}\exp(2i\omega t)}{1+\frac{1}{n^2}\frac{h}{\kappa}\frac{CL^2}{\pi^2S}+\frac{2i\omega\gamma}{n^2}}\bigg].
\end{align}
The remaining terms give $T_\mathrm{dc}(x)$:
\begin{align}
\label{eq:temp rise with gas}
        \Delta T_\mathrm{dc}(x) = \frac{2 I_0^2 R L}{\pi^3 \kappa S}\sum_{n = 1}^{\infty}\frac{1-(-1)^n}{2 n^3}\sin\frac{n\pi x}{L}\frac{1}{1+\frac{1}{n^2}\frac{h}{\kappa}\frac{CL^2}{\pi^2S}}.
\end{align}

To solve for $V_{3\omega}$, we first calculate the fluctuation of the resistance, $\delta R$, that results from the fluctuation in temperature, $\Delta T(x,t)$, using the relation
\begin{equation}
    \label{eq:resistfluc}
    \delta R = \frac{R'}{L} \int_{0}^L\Delta T(x,t)\,\mathrm{d}x.
\end{equation}
Substituting Equation~(\ref{eq:variation}) into Equation~(\ref{eq:resistfluc}) gives
\begin{eqnarray}
    \label{eq:resistvariation}
    \delta R = \frac{2I_0^2 RR'L}{\pi^3\kappa S}\sum_{n = 1}^{\infty}\frac{\left[1-(-1)^n\right]^2}{2n^4\pi}\nonumber\\ \times\ \left[ \frac{1}{1+\frac{h'\gamma}{n^2}}-\frac{\frac{1}{2}e^{-2i\omega t}}{1+\frac{h'\gamma}{n^2}-\frac{2i\omega \gamma}{n^2}}-\frac{\frac{1}{2}e^{2i\omega t}}{1+\frac{h'\gamma}{n^2}+\frac{2i\omega\gamma}{n^2}}\right].
\end{eqnarray}
Keeping only the $n = 1$ term and multiplying the total resistance, $R + \delta R$, by $I_0\sin(\omega t)$ we obtain the 3$\omega$ voltage across the suspended wire 
\begin{eqnarray}
\label{eq:V_3w with gas again}
    V_{3\omega} \approx -\frac{2I_0^3 LRR'}{\pi ^4 \kappa S} \frac{\sin(3\omega t -\phi')}{\sqrt{\left(1+\frac{h}{\kappa}\frac{CL^2}{\pi^2S}\right)^2+(2\omega \gamma)^2 }}
\end{eqnarray}
as presented in the main text. The above equations recover their vacuum case expressions in the limit $h=0$ as presented by \textcite{Lu2001}.

The above derivation requires that for any integer $n\geq 1$:
\begin{equation}
    \frac{n^2}{\gamma} +\frac{hC}{\rho c_p S}\gg\frac{I_0^2 R'}{\rho c_p LS}.
\end{equation}
For this to hold even in the case $h=0$ and $n=1$, we have the following constraint on the wire dimensions:
\begin{equation}
    \frac{d^2}{L}\gg\frac{4I_0^2 R'}{\pi^3 \kappa}.
\end{equation}
If $h>0$, the constraint on the wire dimensions instead is
\begin{equation}
    \frac{d^2}{L} +\frac{4h}{\kappa\pi^2} Ld\gg\frac{4I_0^2 R'}{\pi^3 \kappa}.
\end{equation}
The wire diameter should also not exceed the thermal wavelength given by $\lambda=\sqrt{{\alpha}/{2\omega}}$ for our 1D heat-transfer description to hold. In other words,
\begin{equation}
d\ll\sqrt{\frac{\kappa}{2\omega\rho c_p}}.
\end{equation}
Since $\gamma\propto L^2$ a longer wire will result in a larger time constant. This imposes a practical constraint on the frequency window in experiment where $\omega\ll1$ Hz becomes inconvenient to measure. A practical constraint on the wire length could therefore be given by requiring $\gamma\ll(1~\mathrm{s})$ as
\begin{equation}
    L^2\ll(1~\mathrm{s})\frac{\pi^2\kappa}{\rho c_p}. 
\end{equation}
Note, that the chosen time of 1~s is an arbitrary but instructive choice. If $h>0$, the above constraint instead reads $\gamma_\mathrm{app}\ll(1~\mathrm{s})$ where $\gamma_\mathrm{app}=\frac{\gamma}{1+\frac{h}{\kappa}\frac{4L^2}{\pi^2d}}$. It follows that
\begin{equation}
    \frac{L^2}{1+\frac{h}{\kappa}\frac{4L^2}{\pi^2d}}\ll(1~\mathrm{s})\frac{\pi^2\kappa}{\rho c_p}.
\end{equation}

\section{Numerical Solution of the Heat-Transfer Equation}

Equation~(\ref{eq:1D}) can also be solved numerically using an implicit finite-difference scheme. To do this, we divide the length of the sample into $J$ discrete segments of equal size. Then, at any particular position $i$ along the sample, $\frac{\partial T(x,t)}{\partial t} \approx \frac{T^{n+1}_{i} - T^{n}_{i}}{\Delta t}$ and $\frac{\partial^{2} T(x,t)}{\partial x^{2}} \approx \frac{T^{n+1}_{i+1} - 2T^{n+1}_{i} + T^{n+1}_{i-1}}{(\Delta x)^{2}}$ where $0\le i\le J$ and $0\le n\le N$ denotes the time step. Plugging this into Equation (\ref{eq:1D}) and enforcing the boundary condition that the ends of the wire remain at constant temperature yields
\begin{align}
\begin{cases}
    \bigg(\frac{\rho c_p}{\Delta t} + \frac{2 \kappa}{(\Delta x)^2} + \frac{hC}{S}\bigg) T^{n+1}_i - \frac{\kappa}{(\Delta x)^2}T^{n+1}_{i-1} - \frac{\kappa}{(\Delta x)^2}T^{n+1}_{i+1} \\ = \frac{\rho c_p}{\Delta t} T^n_i + \frac{hC}{S}T_0 +  \frac{I_0^2 \sin^2(\omega n\Delta t)}{LS}[R+R'(T^n_i-T_0)] \\ \\
    T_0^n = T_J^n = T_0.
\end{cases}
\end{align}

Considering all $i$ values along the sample, this gives a system of equations. Therefore, if we know the temperature values along the sample at time step $n$, we can find the temperature profile at time step $n+1$ using the matrix equation 
\begin{align}
\begin{bmatrix}
1 & 0 & 0 & 0 & \ldots & 0 \\
b & a & b & 0 & \ldots & 0\\
0 & b & a & b & \ldots & 0 \\
\vdots & \vdots & \ddots & \ddots & \ddots & \vdots\\
\vdots & \vdots & 0 & b & a & b\\
0 & 0 & 0 & 0 & 0 & 1 
\end{bmatrix}
\begin{bmatrix}
T^{n+1}_0\\
T^{n+1}_1\\
T^{n+1}_2 \\
\vdots\\
\vdots \\
T^{n+1}_J
\end{bmatrix} = 
\begin{bmatrix}
T_0\\
c\\
c \\
\vdots\\
c \\
T_0
\end{bmatrix}
\end{align}
where 
\begin{align}
\begin{cases}
    a = \frac{\rho c_p}{\Delta t} + \frac{2 \kappa}{(\Delta x)^2} + \frac{hC}{S}\\
    b = - \frac{\kappa}{(\Delta x)^2} \\
    c = \frac{\rho c_p}{\Delta t} T^n_i + \frac{hC}{S}T_0 +  \frac{I_0^2 \sin^2(\omega n\Delta t)}{LS}[R+R'(T^n_i-T_0)]
\end{cases}
\end{align}

Numerical solutions of Equation~(\ref{eq:1D}) using this implicit finite-difference scheme are shown in Fig.~\ref{01}(b) where we computed the maximum temperature rise, $T_0+\Delta T_{\mathrm{dc}}+\Delta T_{\mathrm{ac}}$, at several positions along the wire. The strong agreement of the numerical and analytical solutions provides robust validation of our analytical model. The presented implicit finite difference scheme may be adapted to determine the temperature profile along a wire for more complex current wave forms and/or boundary conditions where an analytical solution may not be easily derived.

\section{Errors due to Radiation}

The 1D heat-transfer equation including radiation can be written as:
\begin{align}
\rho c_p  \frac{\partial T(x,t)}{\partial t}  -\kappa \frac {\partial^2 T(x,t)}{\partial x^2 } +\frac{hC}{S}(T(x,t)-T_0) \nonumber \\+ \frac{\epsilon \sigma C}{S}(T^4(x,t)-T_0^4) = \frac{I_0^2 \sin^2\omega t}{LS}[R+R'(T(x,t)-T_0)]
\end{align}
Simplify the equation using Taylor expansion for small values of $T(x,t)-T_0$, we get
\begin{align}
\rho c_p  \frac{\partial T(x,t)}{\partial t}  -\kappa \frac {\partial^2 T(x,t)}{\partial x^2 } +\frac{C}{S}(h+4\epsilon\sigma T_0^3)(T(x,t)-T_0) \nonumber \\ = \frac{I_0^2 \sin^2\omega t}{LS}[R+R'(T(x,t)-T_0)]
\end{align}
Written that way, our model yields an apparent heat loss to the gas $h_\mathrm{app}=h+4\epsilon\sigma T_0^3$. This renders the determination of the wire’s thermal conductivity $\kappa$ more challenging. Using a reference sample such as our Pt wire, one could determine the pressure dependence of $h_\mathrm{app}$, and then use the result to validate a model for $h_\mathrm{app}(p)$. Once a model for $h_\mathrm{app}(p)$ is established, one could in principle use that model in conjunction with 3$\omega$ measurements at finite gas pressure to extract $\kappa$ for a different wire, provided the model still holds. However, even in the presence of radiation heat loss, one may still extract $\rho c_p$ of the wire. This is apparent in the high-frequency expression of $V_{3\omega}$ [Eq. (\ref{eq:v3wHF})], which does not depend on $h$ [or $h_\mathrm{app}(p)$ in this case].

\section{Heat-Transfer Coefficient $h$}

We consider an infinite wire of diameter $d$ at uniform temperature $T_\mathrm{w}$ immersed in an infinite gaseous medium which far from the wire is at temperature $T_0$. In analogy to the consideration for a spherical particle \cite{Yuen1986}, we subdivide the space around the wire into two concentric regions (see inset of Fig.~\ref{fig:03}(b)): one close to the wire with thickness $l_\mathrm{mfp}$ in which molecules may collide with the particle but not with each other, and a second farther away from the wire in which they do collide with each other and continuum conduction holds. To obtain the steady-state temperature field around the wire, we solve Laplace's equation in cylindrical coordinates
\begin{equation}
    \frac{1}{r} \frac{\mathrm{d}}{\mathrm{d}r}\left(\kappa_\mathrm{gas} r\frac{\mathrm{d}T}{\mathrm{d}r}\right)=0.
\end{equation}
This yields a solution of the form
\begin{equation}
    T(r)=\frac{C_1}{\kappa_\mathrm{gas}}\ln(r)+C_2
\end{equation}
where $C_1$ and $C_2$ are constants. To determine $C_2$, we require $T(r_0) \rightarrow T_0$ for $r_0\gg d/2$. This gives
\begin{equation}
\label{eq:C3}
    T(r)=\frac{C_1}{\kappa_\mathrm{gas}}\ln\left(\frac{r}{r_0}\right)+T_0.
\end{equation}

Consider molecules at the boundary between continuum and noncontinuum regions, i.e., at $r=d/2+l_\mathrm{mfp}$. We write as $\Gamma$ the flux density of molecules moving into the noncontinuum region. Since these molecules are randomly directed, only a fraction $\frac{2}{\pi}\arcsin\left(\frac{d}{d+2l_\mathrm{mfp}}\right)$ will hit the wire. The collision rate of gas molecules with the wire over length $L$ is then $4L\left(\frac{d}{2}+l_\mathrm{mfp}\right)\arcsin\left(\frac{d}{d+2l_\mathrm{mfp}}\right)\Gamma$. The average energy exchanged per collision is $\alpha_\mathrm{w} (c_vm_\mathrm{g}+\frac{k_\mathrm{B}}{2})(T_\mathrm{w}-T_\mathrm{b}) $ where $\alpha_\mathrm{w}$ is the accommodation coefficient of the wire as described in the main text, $c_v=f\cdot k_\mathrm{B}/2$ is the specific heat capacity per individual molecule at constant volume with active degrees of freedom $f$, $m_\mathrm{g}$ is the mass of an individual gas molecule, and $T_\mathrm{b}$ is the temperature at the boundary between the noncontinuum and continuum regions. Therefore, the rate of heat transfer from the wire to the gas is 
\begin{align}
    Q_\mathrm{w}=4L\left(\frac{d}{2}+l_\mathrm{mfp}\right)\arcsin\left(\frac{d}{d+2l_\mathrm{mfp}}\right)n_\mathrm{g}\alpha_\mathrm{w} \sqrt{\frac{k_B T_\mathrm{b}}{2\pi m_\mathrm{g}}} \nonumber\\ \times \left(c_vm_\mathrm{g}+\frac{k_B}{2}\right)(T_\mathrm{w}-T_\mathrm{b})
\end{align}
where $n_\mathrm{g}$ is the number density of gas molecules. We can also determine the heat flow in the continuum region from Fourier's law:
\begin{equation}
    Q_\mathrm{c}= -\kappa_\mathrm{gas}2\pi rL\frac{\mathrm{d}T}{\mathrm{d}r} = -2\pi LC_1.
\end{equation}
Since $Q_\mathrm{c}=Q_\mathrm{w}$, we can solve for $C_1$. Then, evaluating Equation~(\ref{eq:C3}) at $r=d/2+l_\mathrm{mfp}$, we find
\begin{align}
T_\mathrm{b} = T\left(\frac{d}{2}+l_\mathrm{mfp}\right)=\frac{C_1}{\kappa_\mathrm{gas}}\ln\left(\frac{d/2+l_\mathrm{mfp}}{r_0}\right)+T_0\nonumber\\
=\frac{T_0+\frac{2(d+2l_\mathrm{mfp})\arcsin\left(\frac{d}{d+2l_\mathrm{mfp}}\right)\ln\left(\frac{r_0}{d/2+l_\mathrm{mfp}}\right)}{l_\mathrm{mfp}\pi^2}\cdot\alpha_\mathrm{w}\Phi\cdot T_\mathrm{w}}{1+\frac{2(d+2l_\mathrm{mfp})\arcsin\left(\frac{d}{d+2l_\mathrm{mfp}}\right)\ln\left(\frac{r_0}{d/2+l_\mathrm{mfp}}\right)}{l_\mathrm{mfp}\pi^2}\cdot\alpha_\mathrm{w}\Phi}.
\end{align}
where $\Phi= (k_\mathrm{B} + 2c_vm_\mathrm{g})/(4c_vm_\mathrm{g}\cdot C^\ast)$ and $C^\ast$ is defined through the kinetic theory expression $\kappa_\mathrm{gas}=C^\ast\cdot\pi c_vm_\mathrm{g}n_\mathrm{g}l_\mathrm{mfp}\sqrt{k_\mathrm{B}T_\mathrm{b}/2\pi m_\mathrm{g}}$ \cite{Yuen1986}. For a monatomic gas with $c_vm_\mathrm{g}=3k_\mathrm{B}/2$ and $C^\ast=25/16$, this gives $\Phi=32/75$, whereas for a diatomic gas such as nitrogen at room temperature with $c_vm_\mathrm{g}=5k_\mathrm{B}/2$ and $C^\ast=19/16$, this gives $\Phi=48/95$ \cite{Yuen1986}.

Plugging $T_\mathrm{b}$ back into the $Q_\mathrm{w}$ expression yields
\begin{equation}
    Q_\mathrm{w} = \frac{4\kappa_\mathrm{gas}L\alpha_\mathrm{w}\Phi}{\frac{l_\mathrm{mfp}\pi}{(d+2l_\mathrm{mfp})\arcsin\left(\frac{d}{d+2l_\mathrm{mfp}}\right)}+\frac{2}{\pi}\alpha_\mathrm{w}\Phi\ln\left(\frac{r_0}{d/2+l_\mathrm{mfp}}\right)}(T_\mathrm{w}-T_0).
\end{equation}
Finally, using $h=Q_\mathrm{w}/[\pi dL(T_\mathrm{w}-T_0)]$, we obtain
\begin{equation}
\label{eq:hmodel}
    h=\frac{\kappa_\mathrm{gas}}{\frac{dl_\mathrm{mfp}\pi^2}{4\alpha_\mathrm{w}\Phi(d+2l_\mathrm{mfp})\arcsin\left(\frac{d}{d+2l_\mathrm{mfp}}\right)}+\frac{d}{2}\ln\left(\frac{2r_0}{d+2l_\mathrm{mfp}}\right)}.
\end{equation}
For $l_\mathrm{mfp}\gtrsim d/2$, we can use $\arcsin\left(\frac{d}{d+2l_\mathrm{mfp}}\right)\approx\frac{d}{d+2l_\mathrm{mfp}}$ to simplify Eq.~(\ref{eq:hmodel}) as follows:
\begin{equation}
    h\approx \frac{\kappa_\mathrm{gas}}{\frac{l_\mathrm{mfp}\pi^2}{4\alpha_\mathrm{w}\Phi}+\frac{d}{2}\ln\left(\frac{2r_0}{d+2l_\mathrm{mfp}}\right)}.
\end{equation}
In the continuum limit, $l_\mathrm{mfp}\ll d$ and $Kn\rightarrow0$, Equation~(\ref{eq:hmodel}) becomes
\begin{equation}
\label{eq:hmodelcont}
    h=\frac{2\kappa_\mathrm{gas}}{d\ln\left(\frac{2r_0}{d}\right)}.
\end{equation}
This expression immediately follows from the textbook solution $Q_\mathrm{w}$ of Fourier's law applied to steady-state radial thermal conduction in a hollow cylinder with thermal conductivity $\kappa_\mathrm{gas}$, an inner wall of radius $d/2$ at temperature $T_\mathrm{w}$, and an outer wall of radius $r_0$ at temperature $T_0$ \cite{Bergman2017}, when similarly requiring $h=Q_\mathrm{w}/[\pi dL(T_\mathrm{w}-T_0)]$. The same continuum limit is reached by a model proposed for carbon nanotubes \cite{Wei2013b,Wei2013}.

\begin{figure}
\includegraphics[width=0.48\textwidth]{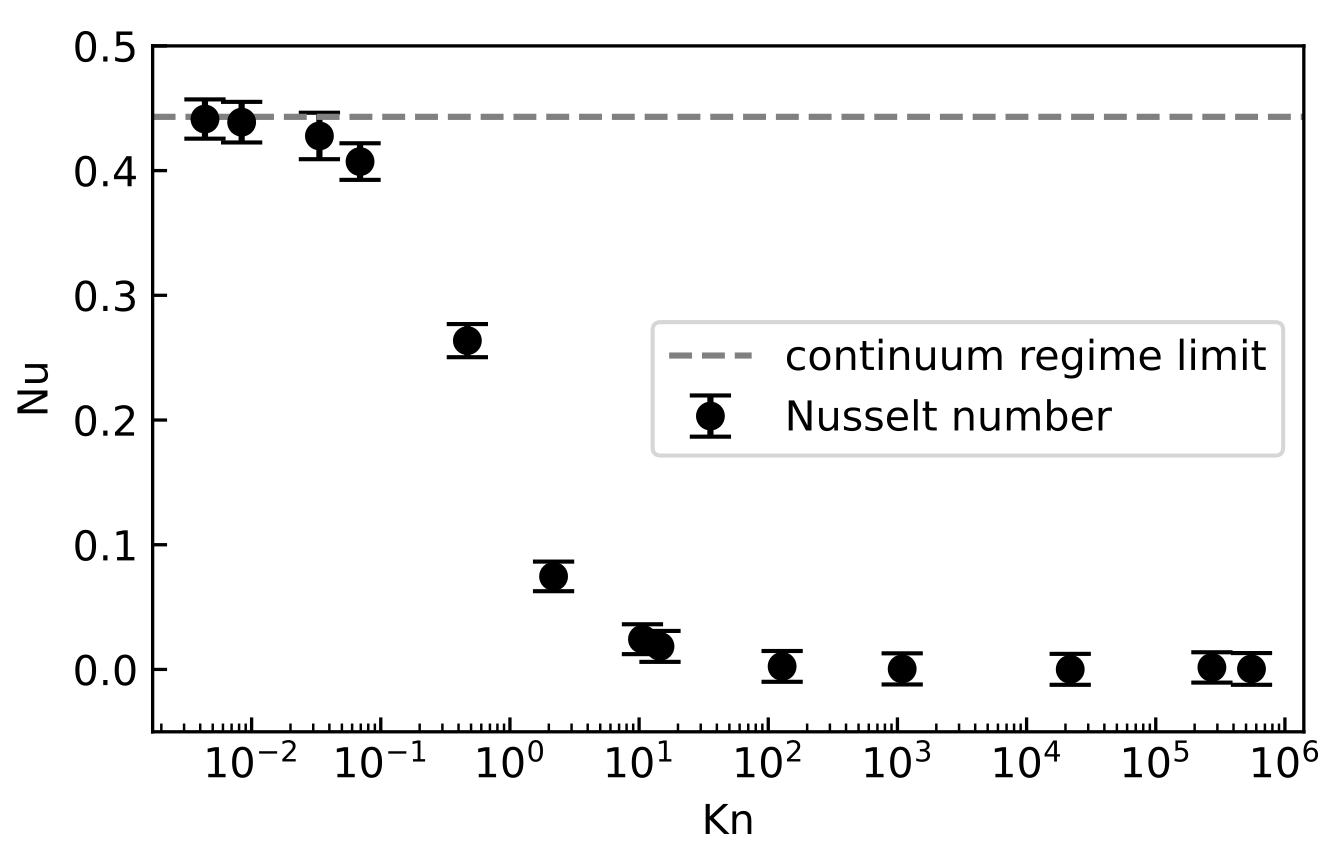}
\caption{\label{fig:A2} Nusselt number $Nu$ as function of Knudsen number $Kn$.}
\end{figure}

While the above model relies on thermal conduction in the gas phase, $h$ extracted from our measurement is agnostic to the nature of the underlying heat transfer (whether it is conductive or convective). To test for indications of convective enhancement, we have calculated the Nusselt number $Nu= {h}/{(\kappa_\mathrm{gas}/d)}$ based on $h$ determined from our experiment. Here we assume $\kappa_\mathrm{gas}=0.026$~W/(m$\cdot$K). The values of $Nu$ thus determined are shown in Fig.~\ref{fig:A2}. These values compare very well with data reported by \textcite{peinado2021free} for free convection heat transfer for Pt wires of similar diameters (12.7 $\mu$m and 25 $\mu$m) measured using a hot wire method in rarified gas atmosphere. In particular, they report $Nu\approx0.37$ at ambient pressure ($Kn\approx3\times10^{-3}$), which is in good agreement with our value of $Nu=0.44\pm0.02$. Our $~20\%$ greater result could be an indication that the actual heat transfer to the gas is slightly enhanced over pure conduction in our experiment.

\begin{thebibliography}{47}%
\makeatletter
\providecommand \@ifxundefined [1]{%
 \@ifx{#1\undefined}
}%
\providecommand \@ifnum [1]{%
 \ifnum #1\expandafter \@firstoftwo
 \else \expandafter \@secondoftwo
 \fi
}%
\providecommand \@ifx [1]{%
 \ifx #1\expandafter \@firstoftwo
 \else \expandafter \@secondoftwo
 \fi
}%
\providecommand \natexlab [1]{#1}%
\providecommand \enquote  [1]{``#1''}%
\providecommand \bibnamefont  [1]{#1}%
\providecommand \bibfnamefont [1]{#1}%
\providecommand \citenamefont [1]{#1}%
\providecommand \href@noop [0]{\@secondoftwo}%
\providecommand \href [0]{\begingroup \@sanitize@url \@href}%
\providecommand \@href[1]{\@@startlink{#1}\@@href}%
\providecommand \@@href[1]{\endgroup#1\@@endlink}%
\providecommand \@sanitize@url [0]{\catcode `\\12\catcode `\$12\catcode `\&12\catcode `\#12\catcode `\^12\catcode `\_12\catcode `\%12\relax}%
\providecommand \@@startlink[1]{}%
\providecommand \@@endlink[0]{}%
\providecommand \url  [0]{\begingroup\@sanitize@url \@url }%
\providecommand \@url [1]{\endgroup\@href {#1}{\urlprefix }}%
\providecommand \urlprefix  [0]{URL }%
\providecommand \Eprint [0]{\href }%
\providecommand \doibase [0]{https://doi.org/}%
\providecommand \selectlanguage [0]{\@gobble}%
\providecommand \bibinfo  [0]{\@secondoftwo}%
\providecommand \bibfield  [0]{\@secondoftwo}%
\providecommand \translation [1]{[#1]}%
\providecommand \BibitemOpen [0]{}%
\providecommand \bibitemStop [0]{}%
\providecommand \bibitemNoStop [0]{.\EOS\space}%
\providecommand \EOS [0]{\spacefactor3000\relax}%
\providecommand \BibitemShut  [1]{\csname bibitem#1\endcsname}%
\let\auto@bib@innerbib\@empty
\bibitem [{\citenamefont {Cahill}\ \emph {et~al.}(2003)\citenamefont {Cahill}, \citenamefont {Ford}, \citenamefont {Goodson}, \citenamefont {Mahan}, \citenamefont {Majumdar}, \citenamefont {Maris}, \citenamefont {Merlin},\ and\ \citenamefont {Phillpot}}]{Cahill2003}%
  \BibitemOpen
  \bibfield  {author} {\bibinfo {author} {\bibfnamefont {D.~G.}\ \bibnamefont {Cahill}}, \bibinfo {author} {\bibfnamefont {W.~K.}\ \bibnamefont {Ford}}, \bibinfo {author} {\bibfnamefont {K.~E.}\ \bibnamefont {Goodson}}, \bibinfo {author} {\bibfnamefont {G.~D.}\ \bibnamefont {Mahan}}, \bibinfo {author} {\bibfnamefont {A.}~\bibnamefont {Majumdar}}, \bibinfo {author} {\bibfnamefont {H.~J.}\ \bibnamefont {Maris}}, \bibinfo {author} {\bibfnamefont {R.}~\bibnamefont {Merlin}},\ and\ \bibinfo {author} {\bibfnamefont {S.~R.}\ \bibnamefont {Phillpot}},\ }\bibfield  {title} {\bibinfo {title} {Nanoscale thermal transport},\ }\href {https://doi.org/10.1063/1.1524305} {\bibfield  {journal} {\bibinfo  {journal} {J. Appl. Phys.}\ }\textbf {\bibinfo {volume} {93}},\ \bibinfo {pages} {793} (\bibinfo {year} {2003})}\BibitemShut {NoStop}%
\bibitem [{\citenamefont {Cahill}\ \emph {et~al.}(2014)\citenamefont {Cahill}, \citenamefont {Braun}, \citenamefont {Chen}, \citenamefont {Clarke}, \citenamefont {Fan}, \citenamefont {Goodson}, \citenamefont {Keblinski}, \citenamefont {King}, \citenamefont {Mahan}, \citenamefont {Majumdar}, \citenamefont {Maris}, \citenamefont {Phillpot}, \citenamefont {Pop},\ and\ \citenamefont {Shi}}]{Cahill2014}%
  \BibitemOpen
  \bibfield  {author} {\bibinfo {author} {\bibfnamefont {D.~G.}\ \bibnamefont {Cahill}}, \bibinfo {author} {\bibfnamefont {P.~V.}\ \bibnamefont {Braun}}, \bibinfo {author} {\bibfnamefont {G.}~\bibnamefont {Chen}}, \bibinfo {author} {\bibfnamefont {D.~R.}\ \bibnamefont {Clarke}}, \bibinfo {author} {\bibfnamefont {S.}~\bibnamefont {Fan}}, \bibinfo {author} {\bibfnamefont {K.~E.}\ \bibnamefont {Goodson}}, \bibinfo {author} {\bibfnamefont {P.}~\bibnamefont {Keblinski}}, \bibinfo {author} {\bibfnamefont {W.~P.}\ \bibnamefont {King}}, \bibinfo {author} {\bibfnamefont {G.~D.}\ \bibnamefont {Mahan}}, \bibinfo {author} {\bibfnamefont {A.}~\bibnamefont {Majumdar}}, \bibinfo {author} {\bibfnamefont {H.~J.}\ \bibnamefont {Maris}}, \bibinfo {author} {\bibfnamefont {S.~R.}\ \bibnamefont {Phillpot}}, \bibinfo {author} {\bibfnamefont {E.}~\bibnamefont {Pop}},\ and\ \bibinfo {author} {\bibfnamefont {L.}~\bibnamefont {Shi}},\ }\bibfield  {title} {\bibinfo {title} {Nanoscale thermal transport. {II}. 2003–2012},\ }\href
  {https://doi.org/10.1063/1.4832615} {\bibfield  {journal} {\bibinfo  {journal} {Appl. Phys. Rev.}\ }\textbf {\bibinfo {volume} {1}},\ \bibinfo {pages} {011305} (\bibinfo {year} {2014})}\BibitemShut {NoStop}%
\bibitem [{\citenamefont {Pop}(2010)}]{Pop2010}%
  \BibitemOpen
  \bibfield  {author} {\bibinfo {author} {\bibfnamefont {E.}~\bibnamefont {Pop}},\ }\bibfield  {title} {\bibinfo {title} {Energy dissipation and transport in nanoscale devices},\ }\href {https://doi.org/10.1007/s12274-010-1019-z} {\bibfield  {journal} {\bibinfo  {journal} {Nano Res.}\ }\textbf {\bibinfo {volume} {3}},\ \bibinfo {pages} {147} (\bibinfo {year} {2010})}\BibitemShut {NoStop}%
\bibitem [{\citenamefont {Balandin}(2011)}]{Balandin2011}%
  \BibitemOpen
  \bibfield  {author} {\bibinfo {author} {\bibfnamefont {A.~A.}\ \bibnamefont {Balandin}},\ }\bibfield  {title} {\bibinfo {title} {Thermal properties of graphene and nanostructured carbon materials},\ }\href@noop {} {\bibfield  {journal} {\bibinfo  {journal} {Nat. Mater.}\ }\textbf {\bibinfo {volume} {10}},\ \bibinfo {pages} {569} (\bibinfo {year} {2011})}\BibitemShut {NoStop}%
\bibitem [{\citenamefont {Cheng}\ \emph {et~al.}(2011)\citenamefont {Cheng}, \citenamefont {Fan}, \citenamefont {Cao}, \citenamefont {Ryu}, \citenamefont {Ji}, \citenamefont {Grigoropoulos},\ and\ \citenamefont {Wu}}]{Cheng2011}%
  \BibitemOpen
  \bibfield  {author} {\bibinfo {author} {\bibfnamefont {C.}~\bibnamefont {Cheng}}, \bibinfo {author} {\bibfnamefont {W.}~\bibnamefont {Fan}}, \bibinfo {author} {\bibfnamefont {J.}~\bibnamefont {Cao}}, \bibinfo {author} {\bibfnamefont {S.~G.}\ \bibnamefont {Ryu}}, \bibinfo {author} {\bibfnamefont {J.}~\bibnamefont {Ji}}, \bibinfo {author} {\bibfnamefont {C.~P.}\ \bibnamefont {Grigoropoulos}},\ and\ \bibinfo {author} {\bibfnamefont {J.}~\bibnamefont {Wu}},\ }\bibfield  {title} {\bibinfo {title} {Heat transfer across the interface between nanoscale solids and gas},\ }\href {https://doi.org/10.1021/nn204072n} {\bibfield  {journal} {\bibinfo  {journal} {ACS Nano}\ }\textbf {\bibinfo {volume} {5}},\ \bibinfo {pages} {10102} (\bibinfo {year} {2011})}\BibitemShut {NoStop}%
\bibitem [{\citenamefont {Corbino}(1910)}]{Corbino1910}%
  \BibitemOpen
  \bibfield  {author} {\bibinfo {author} {\bibfnamefont {O.~M.}\ \bibnamefont {Corbino}},\ }\bibfield  {title} {\bibinfo {title} {Thermische {O}szillationen wechselstromdurchflossener {L}ampen mit dünnem {F}aden und daraus sich ergebende {G}leichrichterwirkung infolge der {A}nwesenheit geradzahliger {O}berschwingungen},\ }\href@noop {} {\bibfield  {journal} {\bibinfo  {journal} {Phys. Z.}\ }\textbf {\bibinfo {volume} {11}},\ \bibinfo {pages} {413} (\bibinfo {year} {1910})}\BibitemShut {NoStop}%
\bibitem [{\citenamefont {Corbino}(1911)}]{Corbino1911}%
  \BibitemOpen
  \bibfield  {author} {\bibinfo {author} {\bibfnamefont {O.~M.}\ \bibnamefont {Corbino}},\ }\bibfield  {title} {\bibinfo {title} {{P}eriodische {W}iderstandsänderungen feiner {M}etallfäden, die durch {W}echselströme zum {G}lühen gebracht werden, sowie {A}bleitung ihrer thermischen {E}igenschaften bei hoher {T}emperatur},\ }\href@noop {} {\bibfield  {journal} {\bibinfo  {journal} {Phys. Z.}\ }\textbf {\bibinfo {volume} {12}},\ \bibinfo {pages} {292} (\bibinfo {year} {1911})}\BibitemShut {NoStop}%
\bibitem [{\citenamefont {Cahill}(1990)}]{Cahill1990}%
  \BibitemOpen
  \bibfield  {author} {\bibinfo {author} {\bibfnamefont {D.~G.}\ \bibnamefont {Cahill}},\ }\bibfield  {title} {\bibinfo {title} {Thermal conductivity measurement from 30 to 750 {K}: the 3$\omega$ method},\ }\href {https://doi.org/10.1063/1.1141498} {\bibfield  {journal} {\bibinfo  {journal} {Rev. Sci. Instrum.}\ }\textbf {\bibinfo {volume} {61}},\ \bibinfo {pages} {802} (\bibinfo {year} {1990})}\BibitemShut {NoStop}%
\bibitem [{\citenamefont {Cahill}\ \emph {et~al.}(1994)\citenamefont {Cahill}, \citenamefont {Katiyar},\ and\ \citenamefont {Abelson}}]{Cahill1994}%
  \BibitemOpen
  \bibfield  {author} {\bibinfo {author} {\bibfnamefont {D.~G.}\ \bibnamefont {Cahill}}, \bibinfo {author} {\bibfnamefont {M.}~\bibnamefont {Katiyar}},\ and\ \bibinfo {author} {\bibfnamefont {J.~R.}\ \bibnamefont {Abelson}},\ }\bibfield  {title} {\bibinfo {title} {Thermal conductivity of a-{S}i:{H} thin films},\ }\href {https://doi.org/10.1103/PhysRevB.50.6077} {\bibfield  {journal} {\bibinfo  {journal} {Phys. Rev. B}\ }\textbf {\bibinfo {volume} {50}},\ \bibinfo {pages} {6077} (\bibinfo {year} {1994})}\BibitemShut {NoStop}%
\bibitem [{\citenamefont {Moon}\ \emph {et~al.}(1996)\citenamefont {Moon}, \citenamefont {Jeong},\ and\ \citenamefont {Kwun}}]{moon19963omega}%
  \BibitemOpen
  \bibfield  {author} {\bibinfo {author} {\bibfnamefont {I.}~\bibnamefont {Moon}}, \bibinfo {author} {\bibfnamefont {Y.~H.}\ \bibnamefont {Jeong}},\ and\ \bibinfo {author} {\bibfnamefont {S.}~\bibnamefont {Kwun}},\ }\bibfield  {title} {\bibinfo {title} {The 3$\omega$ technique for measuring dynamic specific heat and thermal conductivity of a liquid or solid},\ }\href@noop {} {\bibfield  {journal} {\bibinfo  {journal} {Rev. Sci. Instrum.}\ }\textbf {\bibinfo {volume} {67}},\ \bibinfo {pages} {29} (\bibinfo {year} {1996})}\BibitemShut {NoStop}%
\bibitem [{\citenamefont {Lee}\ and\ \citenamefont {Cahill}(1997)}]{Lee1997}%
  \BibitemOpen
  \bibfield  {author} {\bibinfo {author} {\bibfnamefont {S.~M.}\ \bibnamefont {Lee}}\ and\ \bibinfo {author} {\bibfnamefont {D.~G.}\ \bibnamefont {Cahill}},\ }\bibfield  {title} {\bibinfo {title} {Heat transport in thin dielectric films},\ }\href {https://doi.org/10.1063/1.363923} {\bibfield  {journal} {\bibinfo  {journal} {J. Appl. Phys.}\ }\textbf {\bibinfo {volume} {81}},\ \bibinfo {pages} {2590} (\bibinfo {year} {1997})}\BibitemShut {NoStop}%
\bibitem [{\citenamefont {Bhardwaj}\ and\ \citenamefont {Khare}(2022)}]{Bhardwaj2022}%
  \BibitemOpen
  \bibfield  {author} {\bibinfo {author} {\bibfnamefont {R.~G.}\ \bibnamefont {Bhardwaj}}\ and\ \bibinfo {author} {\bibfnamefont {N.}~\bibnamefont {Khare}},\ }\bibfield  {title} {\bibinfo {title} {Review: 3-$\omega$ technique for thermal conductivity measurement—contemporary and advancement in its methodology},\ }\href {https://doi.org/10.1007/s10765-022-03056-3} {\bibfield  {journal} {\bibinfo  {journal} {Int. J. Thermophys.}\ }\textbf {\bibinfo {volume} {43}},\ \bibinfo {pages} {139} (\bibinfo {year} {2022})}\BibitemShut {NoStop}%
\bibitem [{\citenamefont {Lu}\ \emph {et~al.}(2001)\citenamefont {Lu}, \citenamefont {Yi},\ and\ \citenamefont {Zhang}}]{Lu2001}%
  \BibitemOpen
  \bibfield  {author} {\bibinfo {author} {\bibfnamefont {L.}~\bibnamefont {Lu}}, \bibinfo {author} {\bibfnamefont {W.}~\bibnamefont {Yi}},\ and\ \bibinfo {author} {\bibfnamefont {D.~L.}\ \bibnamefont {Zhang}},\ }\bibfield  {title} {\bibinfo {title} {3$\omega$ method for specific heat and thermal conductivity measurements},\ }\href {https://doi.org/10.1063/1.1378340} {\bibfield  {journal} {\bibinfo  {journal} {Rev. Sci. Instrum.}\ }\textbf {\bibinfo {volume} {72}},\ \bibinfo {pages} {2996} (\bibinfo {year} {2001})}\BibitemShut {NoStop}%
\bibitem [{\citenamefont {Dames}\ and\ \citenamefont {Chen}(2005)}]{Dames2005}%
  \BibitemOpen
  \bibfield  {author} {\bibinfo {author} {\bibfnamefont {C.}~\bibnamefont {Dames}}\ and\ \bibinfo {author} {\bibfnamefont {G.}~\bibnamefont {Chen}},\ }\bibfield  {title} {\bibinfo {title} {1$\omega$, 2$\omega$, and 3$\omega$ methods for measurements of thermal properties},\ }\href {https://doi.org/10.1063/1.2130718} {\bibfield  {journal} {\bibinfo  {journal} {Rev. Sci. Instrum.}\ }\textbf {\bibinfo {volume} {76}},\ \bibinfo {pages} {124902} (\bibinfo {year} {2005})}\BibitemShut {NoStop}%
\bibitem [{\citenamefont {Hou}\ \emph {et~al.}(2006)\citenamefont {Hou}, \citenamefont {Wang}, \citenamefont {Vellelacheruvu}, \citenamefont {Guo}, \citenamefont {Liu},\ and\ \citenamefont {Cheng}}]{Hou2006}%
  \BibitemOpen
  \bibfield  {author} {\bibinfo {author} {\bibfnamefont {J.}~\bibnamefont {Hou}}, \bibinfo {author} {\bibfnamefont {X.}~\bibnamefont {Wang}}, \bibinfo {author} {\bibfnamefont {P.}~\bibnamefont {Vellelacheruvu}}, \bibinfo {author} {\bibfnamefont {J.}~\bibnamefont {Guo}}, \bibinfo {author} {\bibfnamefont {C.}~\bibnamefont {Liu}},\ and\ \bibinfo {author} {\bibfnamefont {H.-M.}\ \bibnamefont {Cheng}},\ }\bibfield  {title} {\bibinfo {title} {Thermal characterization of single-wall carbon nanotube bundles using the self-heating 3$\omega$ technique},\ }\href {https://doi.org/10.1063/1.2402973} {\bibfield  {journal} {\bibinfo  {journal} {J. Appl. Phys.}\ }\textbf {\bibinfo {volume} {100}},\ \bibinfo {pages} {124314} (\bibinfo {year} {2006})}\BibitemShut {NoStop}%
\bibitem [{\citenamefont {Tong}\ and\ \citenamefont {Majumdar}(2006)}]{Tong2006}%
  \BibitemOpen
  \bibfield  {author} {\bibinfo {author} {\bibfnamefont {T.}~\bibnamefont {Tong}}\ and\ \bibinfo {author} {\bibfnamefont {A.}~\bibnamefont {Majumdar}},\ }\bibfield  {title} {\bibinfo {title} {Reexamining the 3-omega technique for thin film thermal characterization},\ }\href {https://doi.org/10.1063/1.2349601} {\bibfield  {journal} {\bibinfo  {journal} {Rev. Sci. Instrum.}\ }\textbf {\bibinfo {volume} {77}},\ \bibinfo {pages} {104902} (\bibinfo {year} {2006})}\BibitemShut {NoStop}%
\bibitem [{\citenamefont {Dames}(2013)}]{Dames2013}%
  \BibitemOpen
  \bibfield  {author} {\bibinfo {author} {\bibfnamefont {C.}~\bibnamefont {Dames}},\ }\bibfield  {title} {\bibinfo {title} {Measuring the thermal conductivity of thin films: 3 omega and related electrothermal methods},\ }\href@noop {} {\bibfield  {journal} {\bibinfo  {journal} {Annu. Rev. Heat Transf.}\ }\textbf {\bibinfo {volume} {16}},\ \bibinfo {pages} {7} (\bibinfo {year} {2013})}\BibitemShut {NoStop}%
\bibitem [{\citenamefont {Jaffe}\ \emph {et~al.}(2020)\citenamefont {Jaffe}, \citenamefont {Smith}, \citenamefont {Brar}, \citenamefont {Lagally},\ and\ \citenamefont {Eriksson}}]{Jaffe2020}%
  \BibitemOpen
  \bibfield  {author} {\bibinfo {author} {\bibfnamefont {G.~R.}\ \bibnamefont {Jaffe}}, \bibinfo {author} {\bibfnamefont {K.~J.}\ \bibnamefont {Smith}}, \bibinfo {author} {\bibfnamefont {V.~W.}\ \bibnamefont {Brar}}, \bibinfo {author} {\bibfnamefont {M.~G.}\ \bibnamefont {Lagally}},\ and\ \bibinfo {author} {\bibfnamefont {M.~A.}\ \bibnamefont {Eriksson}},\ }\bibfield  {title} {\bibinfo {title} {Three-omega thermal-conductivity measurements with curved heater geometries},\ }\href {https://doi.org/10.1063/5.0011627} {\bibfield  {journal} {\bibinfo  {journal} {Appl. Phys. Lett.}\ }\textbf {\bibinfo {volume} {117}},\ \bibinfo {pages} {073102} (\bibinfo {year} {2020})}\BibitemShut {NoStop}%
\bibitem [{\citenamefont {Oh}\ \emph {et~al.}(2008)\citenamefont {Oh}, \citenamefont {Jain}, \citenamefont {Eaton}, \citenamefont {Goodson},\ and\ \citenamefont {Lee}}]{oh2008thermal}%
  \BibitemOpen
  \bibfield  {author} {\bibinfo {author} {\bibfnamefont {D.-W.}\ \bibnamefont {Oh}}, \bibinfo {author} {\bibfnamefont {A.}~\bibnamefont {Jain}}, \bibinfo {author} {\bibfnamefont {J.~K.}\ \bibnamefont {Eaton}}, \bibinfo {author} {\bibfnamefont {K.~E.}\ \bibnamefont {Goodson}},\ and\ \bibinfo {author} {\bibfnamefont {J.~S.}\ \bibnamefont {Lee}},\ }\bibfield  {title} {\bibinfo {title} {Thermal conductivity measurement and sedimentation detection of aluminum oxide nanofluids by using the 3$\omega$ method},\ }\href@noop {} {\bibfield  {journal} {\bibinfo  {journal} {Int. J. Heat Fluid Flow}\ }\textbf {\bibinfo {volume} {29}},\ \bibinfo {pages} {1456} (\bibinfo {year} {2008})}\BibitemShut {NoStop}%
\bibitem [{\citenamefont {Lee}(2009)}]{lee2009thermal}%
  \BibitemOpen
  \bibfield  {author} {\bibinfo {author} {\bibfnamefont {S.-M.}\ \bibnamefont {Lee}},\ }\bibfield  {title} {\bibinfo {title} {Thermal conductivity measurement of fluids using the 3$\omega$ method},\ }\href@noop {} {\bibfield  {journal} {\bibinfo  {journal} {Rev. Sci. Instrum.}\ }\textbf {\bibinfo {volume} {80}},\ \bibinfo {pages} {024901} (\bibinfo {year} {2009})}\BibitemShut {NoStop}%
\bibitem [{\citenamefont {Schiffres}\ and\ \citenamefont {Malen}(2011)}]{Schiffres2011}%
  \BibitemOpen
  \bibfield  {author} {\bibinfo {author} {\bibfnamefont {S.~N.}\ \bibnamefont {Schiffres}}\ and\ \bibinfo {author} {\bibfnamefont {J.~A.}\ \bibnamefont {Malen}},\ }\bibfield  {title} {\bibinfo {title} {Improved 3-omega measurement of thermal conductivity in liquid, gases, and powders using a metal-coated optical fiber},\ }\href {https://doi.org/10.1063/1.3593372} {\bibfield  {journal} {\bibinfo  {journal} {Rev. Sci. Instrum.}\ }\textbf {\bibinfo {volume} {82}},\ \bibinfo {pages} {064903} (\bibinfo {year} {2011})}\BibitemShut {NoStop}%
\bibitem [{\citenamefont {Wang}\ and\ \citenamefont {Tang}(2013)}]{Wang2013}%
  \BibitemOpen
  \bibfield  {author} {\bibinfo {author} {\bibfnamefont {Z.~L.}\ \bibnamefont {Wang}}\ and\ \bibinfo {author} {\bibfnamefont {D.~W.}\ \bibnamefont {Tang}},\ }\bibfield  {title} {\bibinfo {title} {Investigation of heat transfer around microwire in air environment using 3$\omega$ method},\ }\href {https://doi.org/https://doi.org/10.1016/j.ijthermalsci.2012.08.002} {\bibfield  {journal} {\bibinfo  {journal} {Int. J. Therm. Sci.}\ }\textbf {\bibinfo {volume} {64}},\ \bibinfo {pages} {145} (\bibinfo {year} {2013})}\BibitemShut {NoStop}%
\bibitem [{\citenamefont {Hamilton}\ and\ \citenamefont {Brisson}(2024)}]{hamilton2024modified}%
  \BibitemOpen
  \bibfield  {author} {\bibinfo {author} {\bibfnamefont {B.}~\bibnamefont {Hamilton}}\ and\ \bibinfo {author} {\bibfnamefont {J.}~\bibnamefont {Brisson}},\ }\bibfield  {title} {\bibinfo {title} {Modified 3$\omega$ conductivity technique for measurements of thermal conductivity in cryogenic fluids},\ }in\ \href@noop {} {\emph {\bibinfo {booktitle} {IOP Conference Series: Materials Science and Engineering}}},\ Vol.\ \bibinfo {volume} {1301}\ (\bibinfo {organization} {IOP Publishing},\ \bibinfo {year} {2024})\ p.\ \bibinfo {pages} {012164}\BibitemShut {NoStop}%
\bibitem [{\citenamefont {Yi}\ \emph {et~al.}(1999)\citenamefont {Yi}, \citenamefont {Lu}, \citenamefont {Dian-lin}, \citenamefont {Pan},\ and\ \citenamefont {Xie}}]{Yi1999}%
  \BibitemOpen
  \bibfield  {author} {\bibinfo {author} {\bibfnamefont {W.}~\bibnamefont {Yi}}, \bibinfo {author} {\bibfnamefont {L.}~\bibnamefont {Lu}}, \bibinfo {author} {\bibfnamefont {Z.}~\bibnamefont {Dian-lin}}, \bibinfo {author} {\bibfnamefont {Z.~W.}\ \bibnamefont {Pan}},\ and\ \bibinfo {author} {\bibfnamefont {S.~S.}\ \bibnamefont {Xie}},\ }\bibfield  {title} {\bibinfo {title} {Linear specific heat of carbon nanotubes},\ }\href {https://doi.org/10.1103/PhysRevB.59.R9015} {\bibfield  {journal} {\bibinfo  {journal} {Phys. Rev. B}\ }\textbf {\bibinfo {volume} {59}},\ \bibinfo {pages} {R9015} (\bibinfo {year} {1999})}\BibitemShut {NoStop}%
\bibitem [{\citenamefont {Choi}\ \emph {et~al.}(2006)\citenamefont {Choi}, \citenamefont {Poulikakos}, \citenamefont {Tharian},\ and\ \citenamefont {Sennhauser}}]{Choi2006}%
  \BibitemOpen
  \bibfield  {author} {\bibinfo {author} {\bibfnamefont {T.-Y.}\ \bibnamefont {Choi}}, \bibinfo {author} {\bibfnamefont {D.}~\bibnamefont {Poulikakos}}, \bibinfo {author} {\bibfnamefont {J.}~\bibnamefont {Tharian}},\ and\ \bibinfo {author} {\bibfnamefont {U.}~\bibnamefont {Sennhauser}},\ }\bibfield  {title} {\bibinfo {title} {Measurement of the thermal conductivity of individual carbon nanotubes by the four-point three-$\omega$ method},\ }\href {https://doi.org/10.1021/nl060331v} {\bibfield  {journal} {\bibinfo  {journal} {Nano Lett.}\ }\textbf {\bibinfo {volume} {6}},\ \bibinfo {pages} {1589} (\bibinfo {year} {2006})}\BibitemShut {NoStop}%
\bibitem [{\citenamefont {Xing}\ \emph {et~al.}(2014{\natexlab{a}})\citenamefont {Xing}, \citenamefont {Jensen}, \citenamefont {Munro}, \citenamefont {White}, \citenamefont {Ban},\ and\ \citenamefont {Chirtoc}}]{Xing2014}%
  \BibitemOpen
  \bibfield  {author} {\bibinfo {author} {\bibfnamefont {C.}~\bibnamefont {Xing}}, \bibinfo {author} {\bibfnamefont {C.}~\bibnamefont {Jensen}}, \bibinfo {author} {\bibfnamefont {T.}~\bibnamefont {Munro}}, \bibinfo {author} {\bibfnamefont {B.}~\bibnamefont {White}}, \bibinfo {author} {\bibfnamefont {H.}~\bibnamefont {Ban}},\ and\ \bibinfo {author} {\bibfnamefont {M.}~\bibnamefont {Chirtoc}},\ }\bibfield  {title} {\bibinfo {title} {Accurate thermal property measurement of fine fibers by the 3-omega technique},\ }\href {https://doi.org/https://doi.org/10.1016/j.applthermaleng.2014.07.035} {\bibfield  {journal} {\bibinfo  {journal} {Appl. Therm. Eng.}\ }\textbf {\bibinfo {volume} {73}},\ \bibinfo {pages} {317} (\bibinfo {year} {2014}{\natexlab{a}})}\BibitemShut {NoStop}%
\bibitem [{\citenamefont {Xing}\ \emph {et~al.}(2014{\natexlab{b}})\citenamefont {Xing}, \citenamefont {Jensen}, \citenamefont {Munro}, \citenamefont {White}, \citenamefont {Ban},\ and\ \citenamefont {Chirtoc}}]{Xing2014b}%
  \BibitemOpen
  \bibfield  {author} {\bibinfo {author} {\bibfnamefont {C.}~\bibnamefont {Xing}}, \bibinfo {author} {\bibfnamefont {C.}~\bibnamefont {Jensen}}, \bibinfo {author} {\bibfnamefont {T.}~\bibnamefont {Munro}}, \bibinfo {author} {\bibfnamefont {B.}~\bibnamefont {White}}, \bibinfo {author} {\bibfnamefont {H.}~\bibnamefont {Ban}},\ and\ \bibinfo {author} {\bibfnamefont {M.}~\bibnamefont {Chirtoc}},\ }\bibfield  {title} {\bibinfo {title} {Thermal property characterization of fine fibers by the 3-omega technique},\ }\href {https://doi.org/https://doi.org/10.1016/j.applthermaleng.2014.06.022} {\bibfield  {journal} {\bibinfo  {journal} {Appl. Therm. Eng.}\ }\textbf {\bibinfo {volume} {71}},\ \bibinfo {pages} {589} (\bibinfo {year} {2014}{\natexlab{b}})}\BibitemShut {NoStop}%
\bibitem [{\citenamefont {Mishra}\ \emph {et~al.}(2020)\citenamefont {Mishra}, \citenamefont {Garnier}, \citenamefont {Le~Corre},\ and\ \citenamefont {Boyard}}]{Mishra2020}%
  \BibitemOpen
  \bibfield  {author} {\bibinfo {author} {\bibfnamefont {K.}~\bibnamefont {Mishra}}, \bibinfo {author} {\bibfnamefont {B.}~\bibnamefont {Garnier}}, \bibinfo {author} {\bibfnamefont {S.}~\bibnamefont {Le~Corre}},\ and\ \bibinfo {author} {\bibfnamefont {N.}~\bibnamefont {Boyard}},\ }\bibfield  {title} {\bibinfo {title} {Accurate measurement of the longitudinal thermal conductivity and volumetric heat capacity of single carbon fibers with the 3$\omega$ method},\ }\href {https://doi.org/10.1007/s10973-019-08568-z} {\bibfield  {journal} {\bibinfo  {journal} {J. Therm. Anal. Calorim.}\ }\textbf {\bibinfo {volume} {139}},\ \bibinfo {pages} {1037} (\bibinfo {year} {2020})}\BibitemShut {NoStop}%
\bibitem [{\citenamefont {Sekimoto}\ \emph {et~al.}(2023)\citenamefont {Sekimoto}, \citenamefont {Abe}, \citenamefont {Kojima}, \citenamefont {Benten},\ and\ \citenamefont {Nakamura}}]{Sekimoto2023}%
  \BibitemOpen
  \bibfield  {author} {\bibinfo {author} {\bibfnamefont {Y.}~\bibnamefont {Sekimoto}}, \bibinfo {author} {\bibfnamefont {R.}~\bibnamefont {Abe}}, \bibinfo {author} {\bibfnamefont {H.}~\bibnamefont {Kojima}}, \bibinfo {author} {\bibfnamefont {H.}~\bibnamefont {Benten}},\ and\ \bibinfo {author} {\bibfnamefont {M.}~\bibnamefont {Nakamura}},\ }\bibfield  {title} {\bibinfo {title} {Error factors in precise thermal conductivity measurement using 3$\omega$ method for wire samples},\ }\href {https://doi.org/10.1007/s10973-022-11892-6} {\bibfield  {journal} {\bibinfo  {journal} {J. Therm. Anal. Calorim.}\ }\textbf {\bibinfo {volume} {148}},\ \bibinfo {pages} {2285} (\bibinfo {year} {2023})}\BibitemShut {NoStop}%
\bibitem [{\citenamefont {Powell}\ \emph {et~al.}(1966)\citenamefont {Powell}, \citenamefont {Ho},\ and\ \citenamefont {Liley}}]{Powell1966}%
  \BibitemOpen
  \bibfield  {author} {\bibinfo {author} {\bibfnamefont {R.~W.}\ \bibnamefont {Powell}}, \bibinfo {author} {\bibfnamefont {C.~Y.}\ \bibnamefont {Ho}},\ and\ \bibinfo {author} {\bibfnamefont {P.~E.}\ \bibnamefont {Liley}},\ }\bibinfo {title} {Thermal conductivity of selected materials},\ in\ \href@noop {} {\emph {\bibinfo {booktitle} {National Standard Reference Data Series}}},\ Vol.~\bibinfo {volume} {8}\ (\bibinfo  {publisher} {National Bureau of Standards},\ \bibinfo {address} {Washington, DC, USA},\ \bibinfo {year} {1966})\BibitemShut {NoStop}%
\bibitem [{\citenamefont {Furukawa}\ \emph {et~al.}(1974)\citenamefont {Furukawa}, \citenamefont {Reilly},\ and\ \citenamefont {Gallagher}}]{Furukawa1974}%
  \BibitemOpen
  \bibfield  {author} {\bibinfo {author} {\bibfnamefont {G.~T.}\ \bibnamefont {Furukawa}}, \bibinfo {author} {\bibfnamefont {M.~L.}\ \bibnamefont {Reilly}},\ and\ \bibinfo {author} {\bibfnamefont {J.~S.}\ \bibnamefont {Gallagher}},\ }\bibfield  {title} {\bibinfo {title} {Critical analysis of heat—capacity data and evaluation of thermodynamic properties of ruthenium, rhodium, palladium, iridium, and platinum from 0 to 300{K}. {A} survey of the literature data on osmium},\ }\href {https://doi.org/10.1063/1.3253137} {\bibfield  {journal} {\bibinfo  {journal} {J. Phys. Chem. Ref. Data}\ }\textbf {\bibinfo {volume} {3}},\ \bibinfo {pages} {163} (\bibinfo {year} {1974})}\BibitemShut {NoStop}%
\bibitem [{\citenamefont {Springer}(1971)}]{Springer1971}%
  \BibitemOpen
  \bibfield  {author} {\bibinfo {author} {\bibfnamefont {G.~S.}\ \bibnamefont {Springer}},\ }\bibinfo {title} {Heat transfer in rarified gases},\ in\ \href@noop {} {\emph {\bibinfo {booktitle} {Advances in Heat Transfer}}},\ Vol.~\bibinfo {volume} {7},\ \bibinfo {editor} {edited by\ \bibinfo {editor} {\bibfnamefont {J.}~\bibnamefont {T.~F.~Irvine}}\ and\ \bibinfo {editor} {\bibfnamefont {J.~P.}\ \bibnamefont {Hartnett}}}\ (\bibinfo  {publisher} {Academic Press, Inc.},\ \bibinfo {address} {New York, NY, USA},\ \bibinfo {year} {1971})\ pp.\ \bibinfo {pages} {163--218}\BibitemShut {NoStop}%
\bibitem [{\citenamefont {Yuen}\ \emph {et~al.}(1986)\citenamefont {Yuen}, \citenamefont {Miller},\ and\ \citenamefont {Hunt}}]{Yuen1986}%
  \BibitemOpen
  \bibfield  {author} {\bibinfo {author} {\bibfnamefont {W.~W.}\ \bibnamefont {Yuen}}, \bibinfo {author} {\bibfnamefont {F.~J.}\ \bibnamefont {Miller}},\ and\ \bibinfo {author} {\bibfnamefont {A.~J.}\ \bibnamefont {Hunt}},\ }\bibfield  {title} {\bibinfo {title} {Heat transfer characteristics of a gas-particle mixture under direct radiant heating},\ }\href {https://doi.org/https://doi.org/10.1016/0735-1933(86)90054-0} {\bibfield  {journal} {\bibinfo  {journal} {Int. Commun. Heat Mass Transf.}\ }\textbf {\bibinfo {volume} {13}},\ \bibinfo {pages} {145} (\bibinfo {year} {1986})}\BibitemShut {NoStop}%
\bibitem [{\citenamefont {Huber}\ and\ \citenamefont {Harvey}(2011)}]{huber2011thermal}%
  \BibitemOpen
  \bibfield  {author} {\bibinfo {author} {\bibfnamefont {M.~L.}\ \bibnamefont {Huber}}\ and\ \bibinfo {author} {\bibfnamefont {A.~H.}\ \bibnamefont {Harvey}},\ }\bibinfo {title} {Thermal conductivity of gases},\ in\ \href@noop {} {\emph {\bibinfo {booktitle} {CRC Handbook of Chemistry and Physics}}},\ \bibinfo {editor} {edited by\ \bibinfo {editor} {\bibfnamefont {W.~M.}\ \bibnamefont {Haynes}}}\ (\bibinfo  {publisher} {CRC Press},\ \bibinfo {address} {Boca Raton, FL, USA},\ \bibinfo {year} {2011})\ pp.\ \bibinfo {pages} {240--241},\ \bibinfo {edition} {92nd}\ ed.\BibitemShut {Stop}%
\bibitem [{\citenamefont {Dayton}(1998)}]{Dayton1998}%
  \BibitemOpen
  \bibfield  {author} {\bibinfo {author} {\bibfnamefont {B.~B.}\ \bibnamefont {Dayton}},\ }\bibinfo {title} {Kinetic theory of gases},\ in\ \href@noop {} {\emph {\bibinfo {booktitle} {Foundations of Vacuum Science and Technology}}},\ \bibinfo {editor} {edited by\ \bibinfo {editor} {\bibfnamefont {J.~M.}\ \bibnamefont {Lafferty}}}\ (\bibinfo  {publisher} {John Wiley \& Sons, Inc.},\ \bibinfo {address} {New York, NY, USA},\ \bibinfo {year} {1998})\ pp.\ \bibinfo {pages} {1--80}\BibitemShut {NoStop}%
\bibitem [{\citenamefont {Bird}(1994)}]{Bird1994}%
  \BibitemOpen
  \bibfield  {author} {\bibinfo {author} {\bibfnamefont {G.~A.}\ \bibnamefont {Bird}},\ }\href@noop {} {\emph {\bibinfo {title} {Molecular Gas Dynamics And The Direct Simulation Of Gas Flows, Oxford Engineering Science Series}},\ Vol.~{\bibinfo {volume} {42}}}\ (\bibinfo  {publisher} {Oxford University Press, Oxford, United Kingdom},\ \bibinfo {year} {1994})\BibitemShut {NoStop}%
\bibitem [{\citenamefont {Tantos}\ \emph {et~al.}(2015)\citenamefont {Tantos}, \citenamefont {Valougeorgis},\ and\ \citenamefont {Frezzotti}}]{Tantos2015}%
  \BibitemOpen
  \bibfield  {author} {\bibinfo {author} {\bibfnamefont {C.}~\bibnamefont {Tantos}}, \bibinfo {author} {\bibfnamefont {D.}~\bibnamefont {Valougeorgis}},\ and\ \bibinfo {author} {\bibfnamefont {A.}~\bibnamefont {Frezzotti}},\ }\bibfield  {title} {\bibinfo {title} {Conductive heat transfer in rarefied polyatomic gases confined between parallel plates via various kinetic models and the {DSMC} method},\ }\href {https://doi.org/https://doi.org/10.1016/j.ijheatmasstransfer.2015.04.092} {\bibfield  {journal} {\bibinfo  {journal} {Int. J. Heat Mass Transf.}\ }\textbf {\bibinfo {volume} {88}},\ \bibinfo {pages} {636} (\bibinfo {year} {2015})}\BibitemShut {NoStop}%
\bibitem [{\citenamefont {Chen}\ \emph {et~al.}(2018)\citenamefont {Chen}, \citenamefont {Wang}, \citenamefont {Yang}, \citenamefont {Li},\ and\ \citenamefont {Zhang}}]{chen2018rough}%
  \BibitemOpen
  \bibfield  {author} {\bibinfo {author} {\bibfnamefont {H.}~\bibnamefont {Chen}}, \bibinfo {author} {\bibfnamefont {H.}~\bibnamefont {Wang}}, \bibinfo {author} {\bibfnamefont {Y.}~\bibnamefont {Yang}}, \bibinfo {author} {\bibfnamefont {N.}~\bibnamefont {Li}},\ and\ \bibinfo {author} {\bibfnamefont {L.}~\bibnamefont {Zhang}},\ }\bibfield  {title} {\bibinfo {title} {Rough boundary effect in thermal transport: A {L}orentz gas model},\ }\href@noop {} {\bibfield  {journal} {\bibinfo  {journal} {Phys. Rev. E}\ }\textbf {\bibinfo {volume} {98}},\ \bibinfo {pages} {032131} (\bibinfo {year} {2018})}\BibitemShut {NoStop}%
\bibitem [{\citenamefont {Wang}\ \emph {et~al.}(2019)\citenamefont {Wang}, \citenamefont {Yang}, \citenamefont {Chen}, \citenamefont {Li},\ and\ \citenamefont {Zhang}}]{wang2019thermal}%
  \BibitemOpen
  \bibfield  {author} {\bibinfo {author} {\bibfnamefont {H.}~\bibnamefont {Wang}}, \bibinfo {author} {\bibfnamefont {Y.}~\bibnamefont {Yang}}, \bibinfo {author} {\bibfnamefont {H.}~\bibnamefont {Chen}}, \bibinfo {author} {\bibfnamefont {N.}~\bibnamefont {Li}},\ and\ \bibinfo {author} {\bibfnamefont {L.}~\bibnamefont {Zhang}},\ }\bibfield  {title} {\bibinfo {title} {Thermal rectification induced by geometrical asymmetry: A two-dimensional multiparticle {L}orentz gas model},\ }\href@noop {} {\bibfield  {journal} {\bibinfo  {journal} {Phys. Rev. E}\ }\textbf {\bibinfo {volume} {99}},\ \bibinfo {pages} {062111} (\bibinfo {year} {2019})}\BibitemShut {NoStop}%
\bibitem [{\citenamefont {Wu}\ \emph {et~al.}(2021)\citenamefont {Wu}, \citenamefont {Yang}, \citenamefont {Lu}, \citenamefont {Wang}, \citenamefont {Xu}, \citenamefont {Yu},\ and\ \citenamefont {Zhang}}]{PhysRevE.103.052135}%
  \BibitemOpen
  \bibfield  {author} {\bibinfo {author} {\bibfnamefont {Y.}~\bibnamefont {Wu}}, \bibinfo {author} {\bibfnamefont {Y.}~\bibnamefont {Yang}}, \bibinfo {author} {\bibfnamefont {L.}~\bibnamefont {Lu}}, \bibinfo {author} {\bibfnamefont {T.}~\bibnamefont {Wang}}, \bibinfo {author} {\bibfnamefont {L.}~\bibnamefont {Xu}}, \bibinfo {author} {\bibfnamefont {Z.}~\bibnamefont {Yu}},\ and\ \bibinfo {author} {\bibfnamefont {L.}~\bibnamefont {Zhang}},\ }\bibfield  {title} {\bibinfo {title} {Ballistic thermal rectification in asymmetric homojunctions},\ }\href {https://doi.org/10.1103/PhysRevE.103.052135} {\bibfield  {journal} {\bibinfo  {journal} {Phys. Rev. E}\ }\textbf {\bibinfo {volume} {103}},\ \bibinfo {pages} {052135} (\bibinfo {year} {2021})}\BibitemShut {NoStop}%
\bibitem [{\citenamefont {Hu}\ \emph {et~al.}(2007)\citenamefont {Hu}, \citenamefont {Shenogin}, \citenamefont {Keblinski},\ and\ \citenamefont {Raravikar}}]{Hu2007}%
  \BibitemOpen
  \bibfield  {author} {\bibinfo {author} {\bibfnamefont {M.}~\bibnamefont {Hu}}, \bibinfo {author} {\bibfnamefont {S.}~\bibnamefont {Shenogin}}, \bibinfo {author} {\bibfnamefont {P.}~\bibnamefont {Keblinski}},\ and\ \bibinfo {author} {\bibfnamefont {N.}~\bibnamefont {Raravikar}},\ }\bibfield  {title} {\bibinfo {title} {Thermal energy exchange between carbon nanotube and air},\ }\href {https://doi.org/10.1063/1.2746954} {\bibfield  {journal} {\bibinfo  {journal} {Appl. Phys. Lett.}\ }\textbf {\bibinfo {volume} {90}},\ \bibinfo {pages} {231905} (\bibinfo {year} {2007})}\BibitemShut {NoStop}%
\bibitem [{\citenamefont {Hsu}\ \emph {et~al.}(2011)\citenamefont {Hsu}, \citenamefont {Pettes}, \citenamefont {Aykol}, \citenamefont {Chang}, \citenamefont {Hung}, \citenamefont {Theiss}, \citenamefont {Shi},\ and\ \citenamefont {Cronin}}]{Hsu2011}%
  \BibitemOpen
  \bibfield  {author} {\bibinfo {author} {\bibfnamefont {I.~K.}\ \bibnamefont {Hsu}}, \bibinfo {author} {\bibfnamefont {M.~T.}\ \bibnamefont {Pettes}}, \bibinfo {author} {\bibfnamefont {M.}~\bibnamefont {Aykol}}, \bibinfo {author} {\bibfnamefont {C.-C.}\ \bibnamefont {Chang}}, \bibinfo {author} {\bibfnamefont {W.-H.}\ \bibnamefont {Hung}}, \bibinfo {author} {\bibfnamefont {J.}~\bibnamefont {Theiss}}, \bibinfo {author} {\bibfnamefont {L.}~\bibnamefont {Shi}},\ and\ \bibinfo {author} {\bibfnamefont {S.~B.}\ \bibnamefont {Cronin}},\ }\bibfield  {title} {\bibinfo {title} {Direct observation of heat dissipation in individual suspended carbon nanotubes using a two-laser technique},\ }\href {https://doi.org/10.1063/1.3627236} {\bibfield  {journal} {\bibinfo  {journal} {J. Appl. Phys.}\ }\textbf {\bibinfo {volume} {110}},\ \bibinfo {pages} {044328} (\bibinfo {year} {2011})}\BibitemShut {NoStop}%
\bibitem [{\citenamefont {Peng}\ \emph {et~al.}(2025)\citenamefont {Peng}, \citenamefont {Ginzburg}, \citenamefont {Dickman}, \citenamefont {Bair},\ and\ \citenamefont {Kuehne}}]{peng_2025_15586066}%
  \BibitemOpen
  \bibfield  {author} {\bibinfo {author} {\bibfnamefont {C.}~\bibnamefont {Peng}}, \bibinfo {author} {\bibfnamefont {J.}~\bibnamefont {Ginzburg}}, \bibinfo {author} {\bibfnamefont {U.}~\bibnamefont {Dickman}}, \bibinfo {author} {\bibfnamefont {J.}~\bibnamefont {Bair}},\ and\ \bibinfo {author} {\bibfnamefont {M.}~\bibnamefont {Kuehne}},\ }\href {https://doi.org/10.5281/zenodo.15586066} {\bibinfo {title} {3$\omega$ thermal characterization of suspended fine wires across continuum to free-molecular gas regimes}} (\bibinfo {year} {2025}),\ \bibinfo {note} {https://doi.org/10.5281/zenodo.15586066}\BibitemShut {NoStop}%
\bibitem [{\citenamefont {Bergman}\ and\ \citenamefont {Lavine}(2017)}]{Bergman2017}%
  \BibitemOpen
  \bibfield  {author} {\bibinfo {author} {\bibfnamefont {T.~L.}\ \bibnamefont {Bergman}}\ and\ \bibinfo {author} {\bibfnamefont {A.~S.}\ \bibnamefont {Lavine}},\ }\href@noop {} {\emph {\bibinfo {title} {Fundamentals of Heat and Mass Transfer}}},\ \bibinfo {edition} {8th}\ ed.\ (\bibinfo  {publisher} {John Wiley \& Sons, Inc.},\ \bibinfo {address} {New York, NY, USA},\ \bibinfo {year} {2017})\BibitemShut {NoStop}%
\bibitem [{\citenamefont {Wang}\ \emph {et~al.}(2013{\natexlab{a}})\citenamefont {Wang}, \citenamefont {Liu}, \citenamefont {Zhang}, \citenamefont {Li}, \citenamefont {Zhang},\ and\ \citenamefont {Wei}}]{Wei2013b}%
  \BibitemOpen
  \bibfield  {author} {\bibinfo {author} {\bibfnamefont {H.-D.}\ \bibnamefont {Wang}}, \bibinfo {author} {\bibfnamefont {J.-H.}\ \bibnamefont {Liu}}, \bibinfo {author} {\bibfnamefont {X.}~\bibnamefont {Zhang}}, \bibinfo {author} {\bibfnamefont {T.-Y.}\ \bibnamefont {Li}}, \bibinfo {author} {\bibfnamefont {R.-F.}\ \bibnamefont {Zhang}},\ and\ \bibinfo {author} {\bibfnamefont {F.}~\bibnamefont {Wei}},\ }\bibfield  {title} {\bibinfo {title} {Heat transfer between an individual carbon nanotube and gas environment in a wide {K}nudsen number regime},\ }\href {https://doi.org/10.1155/2013/181543} {\bibfield  {journal} {\bibinfo  {journal} {J. Nanomater.}\ }\textbf {\bibinfo {volume} {2013}},\ \bibinfo {pages} {181543} (\bibinfo {year} {2013}{\natexlab{a}})}\BibitemShut {NoStop}%
\bibitem [{\citenamefont {Wang}\ \emph {et~al.}(2013{\natexlab{b}})\citenamefont {Wang}, \citenamefont {Liu}, \citenamefont {Guo}, \citenamefont {Zhang}, \citenamefont {Zhang}, \citenamefont {Wei},\ and\ \citenamefont {Li}}]{Wei2013}%
  \BibitemOpen
  \bibfield  {author} {\bibinfo {author} {\bibfnamefont {H.-D.}\ \bibnamefont {Wang}}, \bibinfo {author} {\bibfnamefont {J.-H.}\ \bibnamefont {Liu}}, \bibinfo {author} {\bibfnamefont {Z.-Y.}\ \bibnamefont {Guo}}, \bibinfo {author} {\bibfnamefont {X.}~\bibnamefont {Zhang}}, \bibinfo {author} {\bibfnamefont {R.-F.}\ \bibnamefont {Zhang}}, \bibinfo {author} {\bibfnamefont {F.}~\bibnamefont {Wei}},\ and\ \bibinfo {author} {\bibfnamefont {T.-Y.}\ \bibnamefont {Li}},\ }\bibfield  {title} {\bibinfo {title} {Thermal transport across the interface between a suspended single-walled carbon nanotube and air},\ }\href {https://doi.org/10.1080/15567265.2013.794438} {\bibfield  {journal} {\bibinfo  {journal} {Nanoscale Microscale Thermophys. Eng.}\ }\textbf {\bibinfo {volume} {17}},\ \bibinfo {pages} {349} (\bibinfo {year} {2013}{\natexlab{b}})}\BibitemShut {NoStop}%
\bibitem [{\citenamefont {Peinado}\ \emph {et~al.}(2021)\citenamefont {Peinado}, \citenamefont {Muntean},\ and\ \citenamefont {Perez-Grande}}]{peinado2021free}%
  \BibitemOpen
  \bibfield  {author} {\bibinfo {author} {\bibfnamefont {L.}~\bibnamefont {Peinado}}, \bibinfo {author} {\bibfnamefont {V.}~\bibnamefont {Muntean}},\ and\ \bibinfo {author} {\bibfnamefont {I.}~\bibnamefont {Perez-Grande}},\ }\bibfield  {title} {\bibinfo {title} {A free convection heat transfer correlation for very thin horizontal wires in rarefied atmospheres},\ }\href@noop {} {\bibfield  {journal} {\bibinfo  {journal} {Exp. Therm. Fluid Sci.}\ }\textbf {\bibinfo {volume} {122}},\ \bibinfo {pages} {110295} (\bibinfo {year} {2021})}\BibitemShut {NoStop}%
\end{thebibliography}

%

\end{document}